\documentclass[12pt]{article}

\usepackage{pb-diagram}
\usepackage{latexsym}
\usepackage{amsfonts}
\usepackage[usenames,dvipsnames]{xcolor}
\usepackage{graphicx,amssymb,amsmath,epsfig}
\usepackage[english]{babel}
\usepackage{graphicx}
\usepackage{dcolumn}
\usepackage{bm}

\definecolor{myblue}{rgb}{0,0,0.8}
\usepackage[colorlinks=true
,urlcolor=myblue	
,anchorcolor=myblue
,citecolor=myblue
,filecolor=myblue
,linkcolor=myblue
,menucolor=myblue
,pagecolor=myblue
,linktocpage=true  
]{hyperref}

\catcode`\@=11
\def\marginnote#1{}

\newcount\hour
\newcount\minute
\newtoks\amorpm
\hour=\time\divide\hour by60 \minute=\time{\multiply\hour by60
\global\advance\minute by-\hour}\edef\standardtime{{\ifnum\hour<12
\global\amorpm={am}%
        \else\global\amorpm={pm}\advance\hour by-12 \fi
        \ifnum\hour=0 \hour=12 \fi
        \number\hour:\ifnum\minute<10
0\fi\number\minute\the\amorpm}}
\edef\militarytime{\number\hour:\ifnum\minute<10 0\fi\number\minute}

\def\draftlabel#1{{\@bsphack\if@filesw {\let\thepage\relax
   \xdef\@gtempa{\write\@auxout{\string
      \newlabel{#1}{{\@currentlabel}{\thepage}}}}}\@gtempa
   \if@nobreak \ifvmode\nobreak\fi\fi\fi\@esphack}
        \gdef\@eqnlabel{#1}}
\def\@eqnlabel{}
\def\@vacuum{}
\def\draftmarginnote#1{\marginpar{\raggedright\scriptsize\tt#1}}
\def\draft{\oddsidemargin -.5truein
        \def\@oddfoot{\sl preliminary draft \hfil
        \rm\thepage\hfil\sl\today\quad\militarytime}
        \let\@evenfoot\@oddfoot \overfullrule 3pt
        \let\label=\draftlabel
        \let\marginnote=\draftmarginnote

\def\@eqnnum{(\theequation)\rlap{\kern\marginparsep\tt\@eqnlabel}%
\global\let\@eqnlabel\@vacuum}  }

\def\numberbysection{\@addtoreset{equation}{section}
        \def\theequation{\thesection.\arabic{equation}}}

\def\underline#1{\relax\ifmmode\@@underline#1\else
 $\@@underline{\hbox{#1}}$\relax\fi}

\catcode`@=12 \relax

\numberbysection

\topmargin by -\headsep
\def\br{\begin{eqnarray}}
\def\er{\end{eqnarray}}

\def\({\left(}
\def\){\right)}
\def\[{\left[}
\def\]{\right]}

\def\a{\alpha}
\def\b{\beta}

\def\d{\delta}

\def\bpsi{\bar{\psi}}

\def\eps{\epsilon}

\def\g{\gamma}

\def\l{\lambda}
\def\L{\Lambda}

\def\o{\over}
\def\om{\omega}

\def\p{\phi}

\def\pa{\partial}

\def\s{\sigma}

\def\t{\tau}
\def\th{\theta}

\def\tf{\tilde{f}}
\def\tb{\tilde{b}}

\def\bp{{\bar \p}}

\def\vep{\varepsilon}
\def\bvep{\bar{\varepsilon}}

\def\cK{{\cal K}}
\def\si{\sqrt{i}}
\def\sl{\sqrt{\lambda}}
\def\bq{\bar{Q}}

\def\non{\nonumber}
\def\ba{\begin{align}}
\def\ea{\end{align}}
\def\be{\begin{eqnarray}}
\def\ee{\end{eqnarray}}
\def\L{\Lambda}
\def\a{\alpha}
\def\b{\beta}
\def\g{\gamma}
\def\d{\delta}

\def\s{\sigma}
\def\bp{\bar{\psi}}
\def\l{\lambda}

\def\ph{\phi}
\def\p{\psi}
\def\pp{\partial}
\def\o{\omega}

\begin{document}


\vspace*{1cm}
\noindent

\vskip 1 cm
\begin{center}
\noindent{\Large\bf  Type-II  super-B\"acklund transformation and integrable defects for the $N=1$ super sinh-Gordon model}
\end{center}
\normalsize
\vskip 1cm

\begin{center}
{A.R. Aguirre}\footnote{\href{mailto:aleroagu@ift.unesp.br}{aleroagu@ift.unesp.br}}, J.F. Gomes\footnote{\href{mailto:jfg@ift.unesp.br}{jfg@ift.unesp.br}},  N.I. Spano\footnote{\href{mailto:natyspano@ift.unesp.br}{natyspano@ift.unesp.br}} and A.H. Zimerman\footnote{\href{mailto:zimerman@ift.unesp.br}{zimerman@ift.unesp.br}}\\[.8cm]

\par \vskip .1in \noindent
{Instituto de F\'isica Te\'orica - IFT/UNESP,\\
Rua Dr. Bento Teobaldo Ferraz, 271, Bloco II,
01140-070,\\ S\~ao Paulo, Brasil.}
\vskip 2cm

\end{center}

\noindent A new super-B\"acklund transformation for the $N=1$ supersymmetric sinh-Gordon equation is constructed. Based on this construction we propose a type-II integrable defect  for the supersymmetric sinh-Gordon model consistent with this new transformation through the Lagrangian formalism.  Explicit expressions for the modified conserved energy, momentum and supercharges are also computed. In addition, we show for the model that the type-II defect can also been regarded as a pair of fused defects of a previously  introduced type. The explicit derivation of the associated defect matrices is also presented as a necessary condition for the integrability of the model.

\vfill


\newpage
\tableofcontents

\vskip .4in

\section{Introduction}
\label{sec:intro}

Recently there has been great progress in the study of the integrable defects in two-dimensio\-nal classical field theories. Defects, as originally introduced in \cite{Bow1, Bow2}, can be understood as internal boundary conditions linking fields of both sides of it, and described by a local Lagrangian density. It was shown for several models in \cite{Bow1}--\cite{Ale} that these defect conditions are related to the B\"acklund transformations frozen at the location of the defect, and preserve integrability of the original bulk theory after including some compensating contributions to the conserved quantities. This kind of defect is named type-I  if the fields on either side of it only interact with each other at the defect. And it is called type-II if they interact through additional degrees of freedom present only at the defect point \cite{Corr4}. This type-II formulation proved to be suitable not only for describing defect within the $a_2^{(2)}$ Toda model, which had been excluded from the type-I setting, but it also provided additional types of defect for the sine-Gordon and other Toda models \cite{Corr5}. Interestingly, it was established a  relationship between these two type of defects. At least  for the sine-Gordon model \cite{Corr4, Corr10}, and in general for $a_r^{(1)}$ affine Toda field theory \cite{CR}, the type-II defects can be regarded as  fused pairs of type-I defects. However, the type-II defects can  be allowed in models that cannot support type-I defects, as it was shown for the Tzitz\'eica or $a_2^{(2)}$ Toda model \cite{Corr4}. 
In fact, it was also shown in \cite{Corr4, CR} that the type-II defect in the  $a_2^{(2)}$ Toda model can be interpreted as being the result of folding a type-II defect in the $a_2^{(1)}$ Toda model.

The question of deriving the associated infinite set of conserved quantities in the presence of defects has been initially handled by using the Lax approach for a wide class of models \cite{Cau1}--\cite{Ale4}. The integrable defect conditions are encoded within a defect matrix, which allows to compute the modified conserved charges for the total system. 

On the other hand, the question of involutivity of the modified conserved charges has been addressed  by using intensively the algebraic framework of the classical $r$-matrix approach \cite{Kundu}--\cite{Doi6}. In this context, the description of the integrable defects requires to introduce a modified transition matrix satisfying an appropriate Poisson algebra.

At first sight, it seemed not to exist a direct way of linking these two different \mbox{approaches}
for integrable defects. However, very recently a new approach to the subject has been proposed to provide a link between the two previous points of view  by using a multisymplectic formalism \cite{Cau2, Cau3}. This approach was successfully implemented for the nonlinear Schr\"odinger (NLS) equation and the sine-Gordon model.

The main purpose of this paper is to propose a supersymmetric extension of the type-II integrable defect for the $N=1$ super sinh-Gordon (sshG) model, and to study the integrability of the system through the Lagrangian formalism and Lax approach. The presence of integrable defects  in the $N=1$ sshG model has been already discussed in \cite{FLZ1, Nathaly1}. However, the kind of defect introduced in those papers can be regarded as a ``partially'' type-II defect since only auxiliary fermionic fields appears in the defect Lagrangian, and consequently it reduces to type-I defect for sinh-Gordon model in the bosonic limit, where the fermionic fields vanish. 

Our idea in this paper is rather to find a direct supersymmetric extension of the \mbox{type-II} defect of the sine-Gordon model proposed in \cite{Corr4}. The program of finding supersymmetric extensions of integrable type-II defects has started with the $N=1$ super-Liouville  model \cite{Ale6}. The key point in deriving the defect Lagrangian was based on a generalization of the super-B\"acklund transformation for the equation, by including a new (chiral) superfield $\L(x,\th)$ in the formulation. Following the same line of reasoning, in section \ref{backlund} we propose a generalisation of the super-B\"acklund for the $N=1$ sshG model by introducing two new superfields in the description, so that the pure bosonic limit reduces to the type-II defect for the sinh-Gordon model. In section \ref{Lagrangian}, we introduce the Lagrangian description for the type-II defect in the sshG model totally consistent with the new super-B\"acklund.
We also present the supersymmetry transformation that leave the total action invariant and derive the modified conserved supercharge, energy and momentum. The bosonic and fermionic limit are discussed as well.

The fusing procedure will be discussed in section \ref{fusing}. It will be shown that the type-II defect for the sshG model derived from consistency with the proposed super-B\"acklund can be obtained by fusing two defects of the kind given in \cite{FLZ1} for the sshG model.  In section \ref{solutions} we analyse the classical behaviour of one-soliton solutions of sshG equation passing through the type-II defect. Section \ref{final} contained some final remarks and comments on future directions which have emerged from this work.

The integrability of the system will be discussed by computing the associated defect matrices which generate the infinite set of the associated modified conserved quantities. The explicit computations will be presented in appendix \ref{appB}.

\section{Type-II super-B\"acklund for $N=1$ sshG equation}
\label{backlund}
In this section we construct a new super-B\"acklund transformation for the $N=1$ supersymmetric sinh-Gordon (sshG) equation. Let us first introduce a bosonic superfield,
\br
\Phi=\phi-i\theta_{1}\,\bp+i\theta_{2}\,\p - i\theta_{1}\theta_{2} F, \label{B1.1}
\er
where  $\phi$ an $F$ are bosonic fields, $\psi$ and $\bpsi$ are fermionic fields, and   $\theta_i$, $i=1,2$,  elements of the Grassmann algebra. Then, the $N=1$ sshG equation can be written in terms of superfields as follows
\be
D_{+}D_{-}\Phi=im \sinh\Phi,\label{B1.2}
\ee
where the superderivatives are given by
\br
D_{+}=-i\partial_{\theta_{1}}+\theta_{1}\partial_{+}, \qquad D_{-}=i\partial_{\theta_{2}}+\theta_{2}\partial_{-}, \qquad D_{\pm}^2 = \mp i\pa_{\pm}, \qquad \{D_+ ,D_-\} =0.
\er
We denote the light-cone coordinates  $x_{\pm} =x \pm t$, and then $\pa_{\pm} = \frac{1}{2}(\pa_x \pm \pa_t)$. Using the form of the superfield (\ref{B1.1}), we can write eq. (\ref{B1.2}) in components,
\be
\pp_{+}\pp_{-}\ph&=&\frac{m^2}{2}\sinh 2\ph-im\bp\p\sinh\ph,\label{B1.4}\\
\pp_{-}\bp&=&-m\p\cosh\ph,\label{B1.5}\\
\pp_{+}\p&=&-m\bp\cosh\ph,\label{B1.6}
\ee
with the auxiliary field given by $F=m\sinh\phi$. Now, we propose the following super-B\"acklund transformation for the sshG equation connecting the two solutions $\Phi_1$ and $\Phi_2$ as follows,\vspace{-0.3cm}
\br
D_{-}(\Phi_{+}-{\Lambda})&=& \omega_{1}\ e^{\frac{{\Lambda}}{2}}\,f\,\cosh\Big(\frac{\Phi_{-}}{2}\Big),\label{B1.7}\\
D_{+}{\Lambda}&=&\omega_{2}\ e^{-\frac{\left(\Phi_{+}-{\Lambda}\right)}{2}}\,\tilde{f}\,\cosh\Big(\frac{\Phi_{-}}{2}\Big),\\
D_{+}\Phi_{-}&=&\om_{3}\ e^{\frac{(\Phi_{+}-{\Lambda})}{2}}\,f
-\om_{2}\ e^{-\frac{\left(\Phi_{+}-{\Lambda}\right)}{2}}\,\tilde{f}\,\sinh\Big(\frac{\Phi_{-}}{2}\Big),\label{B1.9}\\
D_{-}\Phi_{-}&=&\om_1 \ e^{\frac{{\Lambda}}{2}}\,f\, \sinh\Big(\frac{\Phi_{-}}{2}\Big)+\om_{4}\ e^{-\frac{{\Lambda}}{2}}\,\tilde{f},\label{B1.10}
\er
where $\Phi_{\pm}=\Phi_1\pm \Phi_2$, and we have introduced two fermionic superfields  $f$ and $\tilde{f}$, as well as a new bosonic superfield ${\Lambda}$, which satisfy  respectively the following equations,
\begin{eqnarray}
D_{+}f &=&\frac{im}{\om_{1}}\ e^{\frac{(\Phi_{+}-{\Lambda)}}{2}},\qquad \,\,\,
D_{-}f=-\frac{im}{\omega_{3}}\ e^{\frac{{\Lambda}}{2}}\,\sinh\Big(\frac{\Phi_{-}}{2}\Big),\\
D_{-}\tilde{f} &=& \frac{im}{\omega_{2}}\ e^{-\frac{{\Lambda}}{2}}, \qquad \qquad D_{+}\tilde{f}=\frac{im}{\omega_{4}}\ e^{-\frac{\left(\Phi_{+}-{\Lambda}\right)}{2}}\ \sinh\Big(\frac{\Phi_{-}}{2}\Big). \label{B1.12}
\end{eqnarray}
Here, $\{\omega_k\}_{k=1}^{4}$ are four arbitrary constant parameters. By cross-differentiating eqs. (\ref{B1.9}) and (\ref{B1.10}) we find that if $\Phi_1$ satisfies the sshG equation then $\Phi_2$ also satisfies it. Besides introducing a new bosonic superfield $\L$ in eqs. (\ref{B1.7})--(\ref{B1.12}), the supersymmetry and the Grassmannian property of the superderivatives $D_\pm$ require the introduction of not only one fermionic superfield $f$ like previously was proposed in \cite{ChaiKul}, \mbox{but two fermionic superfields $(f,\tf)$.}

\noindent The additional superfield $\tf$ allows the possibility of having independent contributions coming from the terms of the form $\exp(\pm \L/2)$. The introduction of this new superfield might be better understood in section \ref{fusing} by discussing the fusing procedure. Note also that even when  $\L=0$ there is no direct limit between the super-B\"acklund proposed here and the one given in \cite{ChaiKul}, and then they will be regarded  differently. In addition, when $\psi$ and $\bpsi$ vanish, eqs. (\ref{B1.7})--(\ref{B1.12}) will reduce to the type-II B\"acklund transformation for the sinh-Gordon model \cite{Corr4}. For that reason, they will be called type-II super-B\"acklund transformation for the $N=1$ sshG equation. Now, let us introduce $\th$-expansions for the superfields $\L$, $f$,  and $\tf$, namely \vspace{-0.3cm}
\br
 {\Lambda}&=& {\l}_{0}-i\theta_{1}{\l}_{2}+i\theta_{2}{\l}_{1}-i\theta_{1}\theta_{2}{\l}_{3},\label{B1.13}\\
f&=&f_{1}-i\theta_{1}b_{1}+i\theta_{2}b_{2}-i\theta_{1}\theta_{2}f_{2},\label{B1.14}\\
\tilde{f}&=&\tilde{f}_{1}-i\theta_{1}\tb_{1}+i\theta_{2}\tilde{b}_{2}-i\theta_{1}\theta_{2}\,\tilde{f}_{2},\label{B1.15}
\er
described by six bosonic $\{\l_0,\l_3,b_1,b_2,\tb_1, \tb_2\}$, and  six fermionic $\{\l_1,\l_2, f_1,f_2,\tf_1,$ $\tf_2\}$ fields. Then, we can write down the super-B\"acklund transformation (\ref{B1.7})--(\ref{B1.12}) in components following the same procedure  previously implemented in \cite{FLZ1, Nathaly1}. 
%
%
After some algebra the auxiliary fields $\{F_\pm, b_1, b_2, \tb_1, \tb_2, f_2,$ $\tf_2, \l_1, \l_2,\l_3\}$ can be eliminated, and we obtain the following simplified set of equations containing only the auxiliary fields  $\l_0, f_1, \tf_1$, namely
\br
 \pp_{-}(\ph_{+}-{\l}_{0})\!&=&\!-\frac{m\omega_1}{2\omega_3}\, e^{\l_0}\sinh\ph_{-}+ \frac{i\omega_1}{2}\, e^{\frac{\lambda_{0}}{2}}\cosh\Big(\frac{\ph_{-}}{2}\Big)\psi_+f_1 - \frac{i\om_1\om_4}{2}\sinh\Big(\frac{\ph_{-}}{2}\Big)\tf_1 f_{1}, \label{B1.61}\qquad\\
\pa_+\l_0 &=&-\frac{m\omega_2}{2\omega_4}\, e^{-(\ph_+-\l_0)}\sinh\ph_{-}+ \frac{i\omega_2}{2}\, e^{-\frac{(\phi_+-\lambda_{0})}{2}} 
 \cosh\Big(\frac{\ph_{-}}{2}\Big){\bpsi}_+\tf_1 \non\\&&+\frac{i\om_2\om_3}{2}\sinh\Big(\frac{\ph_{-}}{2}\Big)f_1\tf_{1}, \qquad\\
\pa_+\phi_- &=& \frac{m\omega_2}{\omega_4}e^{-{(\ph_{+}-{\lambda}_{0})}}\sinh^2\Big(\frac{\ph_{-}}{2}\Big) -\frac{m\omega_3}{\omega_1} e^{(\ph_{+}-{\lambda}_{0})}  -\frac{i\omega_3}{2} e^{\frac{(\ph_{+}-{\lambda}_{0})}{2}} \bpsi_+f_1\non \\
&& \!\!\!-\frac{i\omega_2}{2} e^{-\frac{(\ph_{+}-{\lambda}_{0})}{2}}\sinh\Big(\frac{\ph_{-}}{2}\Big)\bp_{+}\tf_1  ,\qquad\,\,\, \mbox{}\\
\pa_-\phi_- &=& -\frac{m\omega_1}{\omega_3}e^{\lambda_{0}}\sinh^2\Big(\frac{\ph_{-}}{2}\Big) +\frac{m\om_4}{\om_2}e^{-\l_0}-\frac{i\om_4}{2}\, e^{-\frac{\l_0}{2}}\psi_+\tf_1\non \\&&+\frac{i\omega_1}{2} e^{\frac{{\lambda}_{0}}{2}}\sinh\Big(\frac{\ph_{-}}{2}\Big) \psi_+ f_1,\\
\p_{-}&=&-\omega_{1}\, e^{\frac{{\lambda}_{0}}{2}}\sinh\Big(\frac{\ph_{-}}{2}\Big)f_{1}-\omega_4\, e^{-\frac{{\lambda}_{0}}{2}}\tilde{f}_{1},\\
 \bp_{-}&=& \omega_2 \, e^{-\frac{(\ph_{+}-{\lambda}_{0})}{2}}\sinh\Big(\frac{\ph_{-}}{2}\Big)\tilde{f}_{1}-\omega_3\, e^{\frac{(\ph_{+}-{\lambda}_{0})}{2}}f_{1},\\
\pa_+ f_1 &=& \frac{m}{2\om_1}\left[e^{\frac{(\ph_{+}-{\lambda}_{0})}{2}}\bpsi_++\om_2\cosh\Big(\frac{\phi_{-}}{2}\Big)\tf_1\],\\
 \pp_{-}f_{1}&=&\frac{{m}}{2\omega_{3}}\Big[e^{\frac{{\lambda}_{0}}{2}}\sinh\Big(\frac{\ph_{-}}{2}\Big)\p_{+}-\om_4 \cosh\Big(\frac{\ph_{-}}{2}\Big)\tf_1\Big],\\
\pa_+ \tf_1 &=&-\frac{m}{2\omega_{4}}\left[ e^{-\frac{\left(\ph_{+}-{\lambda}_{0}\right)}{2}}\sinh\Big(\frac{\ph_{-}}{2}\Big)\bp_{+}+\om_3\cosh\Big(\frac{\ph_{-}}{2}\Big)f_1\],\\
\pp_{-}\tilde{f}_{1}&=&\frac{{m}}{2\omega_{2}}\left[e^{-\frac{{\lambda}_{0}}{2}}\psi_+ +\omega_1\cosh\Big(\frac{\phi_-}{2}\Big) f_1\right]. \label{B1.70}
\er
\newpage
\noindent These equations represent the  type-II super-B\"acklund transformation for $N=1$ sshG model in components,  which depend upon four arbitrary parameters $\{\om_k\}_{k=1}^4$.


\section{Type-II defect for $N=1$ sshG model}
\label{Lagrangian}
Here  we introduce a Lagrangian description of type-II defects in the $N=1$ sshG model which satisfies  super-B\"acklund transformations frozen at $x=0$.

\subsection{Lagrangian description}

The Lagrangian density can be written as follows,
\br
 {\mathcal L} = \theta(-x)  {\mathcal L}_1 + \theta(x) {\mathcal L}_2 +\delta(x) \ {\mathcal L}_D, \label{e2.1}
\er
with
\begin{eqnarray}
 {\mathcal L}_p &=&\frac{1}{2}(\partial_x \phi_p)^2 - \frac{1}{2}(\partial_t \phi_p)^2 +   i\psi_p(\partial_x +\partial_t)\psi_p - i\bar{\psi}_p(\partial_x -\partial_t)\bar{\psi}_p + V_p(\phi_p)\non \\ && \!\! + W_p(\phi_p,\psi_p, \bar{\psi}_p),\qquad\mbox{}\label{e2.2}
\er
where  $\phi_p$ are real scalar fields, and $\psi_p,\bpsi_p$ are the components of Majorana spinor fields in the regions $x<0$ ($p=1$) and $x>0$ ($p=2$) respectively. The bulk potentials are given by,
\begin{eqnarray}
 V_p&=& m^2(\cosh(2\phi_p)-1),\qquad W_p = -4im\,\bar{\psi}_p \psi_p \cosh\phi_p.\label{equ2.3}
\er
Now, the defect Lagrangian density ${\cal L}_D$ can be written in the following way\footnote{From now on, we will use the field coordinates $\phi_{\pm}=\phi_1\pm\phi_2, \psi_{\pm}=\psi_1\pm\psi_2$, and $\bpsi_{\pm}=\bpsi_1\pm\bpsi_2$ to describe expressions associated to the defect.},
\br
   {\mathcal L}_D &=&\phi_- \pa_t \l_0 -\frac{1}{2}\phi_-\pa_t\phi_+ +\frac{i}{2}(\bar{\psi}_+\bar{\psi}_- -\psi_+\psi_- ) + if_1\pa_t f_1 +i\tf_1\pa_t\tf_1 +B_0^{(+)}(\phi_+-\l_0,\phi_-)\nonumber \\[0.1cm]
  &&   + B_0^{(-)}(\phi_-,\l_0)+ B_1^{(+)}(\phi_+-\l_0,\phi_-,\bpsi_+,f_1,\tf_1)+B_1^{(-)}(\phi_-,\l_0,\psi_+,f_1,\tf_1). \label{e2.35}
\end{eqnarray}
Here, we also have introduced  the bosonic field $\l_0(t)$ and the fermionic fields $f_1(t)$, and $\tf_1(t)$, which are all associated with the defect itself at $x=0$. The corresponding defect potentials are given by,
\br
 B_0^{(+)} &=& \om_2^2\, e^{-(\phi_+-\l_0)}\sinh^2\Big(\frac{\phi_-}{2}\Big) +\frac{m^2}{\om_1^2}\,e^{(\phi_+-\l_0)}, \qquad \mbox{}\label{e3.15}\\ 
 B_0^{(-)} &=& \om_1^2\, e^{\l_0}\sinh^2\Big(\frac{\phi_-}{2}\Big)+ \frac{m^2}{\om_2^2} e^{-\l_0},\\
 B^{(+)}_1 &=& i\left[\frac{m}{\om_1} \,e^{\frac{(\phi_+-\l_0)}{2}}\bar{\psi}_+f_1-\om_2 \,e^{-\frac{(\phi_+-\l_0)}{2}}\sinh\Big(\frac{\phi_-}{2}\Big)\bar{\psi}_+\tf_{1}-  \frac{m\om_2}{\om_1}\cosh\Big(\frac{\phi_-}{2}\Big) f_1\tf_1\],\qquad \mbox{}\label{e3.17} \\
B^{(-)}_1&=&- i\left[\frac{m}{\om_2}\, e^{-\frac{{\lambda}_{0}}{2}}\psi_+\tf_1 +\omega_{1}\, e^{\frac{{\lambda}_{0}}{2}}\sinh\Big(\frac{\ph_-}{2}\Big)\psi_+f_1 + \frac{m\om_1}{\om_2}\cosh\Big(\frac{\phi_-}{2}\Big) f_1\tf_1\right],\label{e3.18}
\er
where $\{\om_k\}_{k=1}^2$ are constant, and $m$ is the mass parameter. 
In this setup, we can see that the defect potentials (\ref{e3.15})--(\ref{e3.18}) satisfy the following relations,
\begin{eqnarray}
\frac{\pa B_k^{(+)}}{\pa \phi_+} &=& -\frac{\pa B_k^{(+)}}{\pa \l_0},\qquad \frac{\pa B_k^{(-)}}{\pa \phi_+} =0, \qquad k=0,1,\\  \frac{\pa B_1^{(\pm)}}{\pa \psi_-} &=&0, \qquad  \frac{\pa B_1^{(\pm)}}{\pa \bpsi_-} =0,  \qquad \frac{\pa B_1^{(+)}}{\pa \psi_+} =0, \qquad  \frac{\pa B_1^{(-)}}{\pa \bpsi_+} =0. \qquad \mbox{}
\end{eqnarray}
The bulk fields equations are given by,
\begin{eqnarray}\label{mov1}
\partial_{x}^{2}\phi_{p}-\partial_{t}^{2}\phi_{p}&=&2m^2\sinh(2\phi_{p})-4im\,\bar{\psi}_{p}\psi_{p}\sinh\phi_{p},\nonumber\\
(\partial_{x}-\partial_{t})\bar{\psi}_{p}&=&-2m\,\psi_{p}\cosh\phi_{p},\nonumber\\
(\partial_{x}+\partial_{t})\psi_{p} &=&-2m\,\bar{\psi}_{p}\cosh\phi_{p}, \qquad p=1,2.
\end{eqnarray}
Then, the defect conditions at $x=0$ can be written as follows:
\begin{eqnarray}
\pa_x\phi_+ - \pa_t(\phi_+-\l_0) &=&-{e^{\l_0}}\sinh\phi_-\Big[\om_1^2+\om_2^2\, e^{-\phi_+}\Big]    +{im}\left[\frac{\om_1}{\om_2}+\frac{\om_2}{\om_1}\]\sinh\Big(\frac{\ph_1-\ph_2}{2}\Big)f_1\tf_1\non\\
&&+{i\om_1}e^{\l_0 \over 2}	\cosh\Big(\frac{\ph_-}{2}\Big)\psi_+f_1+{i\om_2} e^{-\frac{(\phi_+ -\l_0)}{ 2}}\cosh\Big(\frac{\ph_-}{2}\Big)\bpsi_+\tf_1 , \label{e2.28}\\[0.1cm]
(\pa_x+\pa_t)\phi_- &=&2\om_2^2 e^{-(\phi_+ -\l_0)}\sinh^2\Big(\frac{\phi_-}{2}\Big)-\frac{2m^2}{\om_1^2}e^{(\phi_+ -\l_0)}  -\frac{im}{\om_1} e^{\frac{(\phi_+-\l_0)}{2}}\bpsi_+f_1\non\\
&& -{i\om_2} e^{-\frac{(\phi_+-\l_0)}{2}}\sinh\Big(\frac{\phi_-}{2}\Big)\bpsi_+\tf_1,\\ 
(\pa_x-\pa_t)\phi_-&=& -2\om_1^2 \,e^{\l_0}\sinh^2\Big(\frac{\phi_-}{2}\Big)+\frac{2m^2}{\om_2^2} e^{-\l_0}-\frac{im}{\om_2}e^{-\frac{\l_0}{2}} \psi_+ \tf_1\non \\
&&+{i\om_1}e^{\l_0 \over 2}	\sinh\Big(\frac{\ph_-}{2}\Big)\psi_+f_1, \\
\psi_- &=& -\o_{1}\, e^{\frac{\lambda_{0}}{2}}\sinh\Big(\frac{\ph_-}{2}\Big)f_{1}-\frac{m}{\om_2}\, e^{-\frac{\lambda_{0}}{2}}\tilde{f}_{1},\label{e2.14}\\
	\bpsi_- &=&\o_{2}\, e^{-\frac{\left(\ph_+-\lambda_{0}\right)}{2}}\sinh\Big(\frac{\ph_-}{2}\Big)\tilde{f}_{1}-\frac{m}{\om_1}\, e^{\frac{(\ph_+-\lambda_{0})}{2}}f_{1},\label{e2.15}\\
	\pp_{t}f_{1} &=&\frac{m}{2\o_{1}} e^{\frac{(\ph_+-\l_{0})}{2}}\bpsi_+ -\frac{\om_1}{2} e^{\frac{\l_{0}}{2}}\sinh\Big(\frac{\ph_-}{2}\Big)\psi_+\non\\&& +\frac{m}{2}\left[\frac{\om_1}{\om_2}+\frac{\om_2}{\om_1}\]\cosh\Big(\frac{\ph_-}{2}\Big)\tf_1,\\
	 \pp_{t}\tf_{1} &=& -\frac{{m}}{2\omega_{2}}\, e^{-\frac{{\lambda}_{0}}{2}}\psi_+ -\frac{\om_2}{2}\, e^{-\frac{\left(\ph_+-{\lambda}_{0}\right)}{2}}\sinh\Big(\frac{\ph_-}{2}\Big)\bp_+\non \\&&  -\frac{m}{2}\left[\frac{\om_1}{\om_2}+\frac{\om_2}{\om_1}\]\cosh\Big(\frac{\ph_-}{2}\Big)f_1.\label{e2.34}
\end{eqnarray}
These defect conditions are consistent with the type-II super-B\"acklund tranformation (\ref{B1.61})--(\ref{B1.70}) for the sshG model frozen at $x=0$, providing that relations $\om_1 \om_3 =m$ and $\om_2 \om_4=m$ are satisfied. Note that the number of equations specifying the defect conditions is less than the number of equations describing the type-II super-B\"acklund. In fact, there is one less defect equation for each auxiliary field we have in the B\"acklund transformation\footnote{In general, if there exist $m$ equations describing the B\"acklund transformation for an integrable equation, with $n<m$ auxiliary fields, then the corresponding defect equations should be described by $(m-n)$ equations valid at the defect point.}. This feature distinguishes the type-II  from the type-I defect, which Lagrangian description leads directly to the frozen B\"acklund transformation itself, because there is no auxiliary fields in the  formulation.

Now, if we apply the operators $\pa_{\pm}$ to eqs.(\ref{e2.14}) and (\ref{e2.15}) respectively, and using properly the above defect conditions, we can verify that the respective $x$-derivatives of the auxiliary fields $f_1$ and $\tf_1$ are given by,
\br
\pa_x f_1 &=&  \frac{m}{2\om_1}\,e^{\frac{(\ph_+-{\lambda}_{0})}{2}}\bpsi_+ +\frac{{\om_1}}{2}\, e^{\frac{{\lambda}_{0}}{2}}\sinh\Big(\frac{\ph_-}{2}\Big)\psi_++\frac{m}{2}\left[\frac{\om_2}{\om_1}-\frac{\om_1}{\om_2}\]\cosh\Big(\frac{\phi_-}{2}\Big)\tf_1,\qquad \mbox{} \\[0.1cm]
\pa_x \tf_1 &=&  -\frac{\om_2}{2}\, e^{-\frac{\left(\ph_+-{\lambda}_{0}\right)}{2}}\sinh\Big(\frac{\ph_-}{2}\Big)\bp_+ +\frac{{m}}{2\omega_{2}}\, e^{-\frac{{\lambda}_{0}}{2}}\psi_+ +\frac{m}{2}\left[\frac{\om_1}{\om_2}-\frac{\om_2}{\om_1}\]\cosh\Big(\frac{\ph_-}{2}\Big)f_1. \qquad \mbox{}
\er
%

%

%
%


%


\subsection{Modified conserved quantities}
\label{conserved}
In this subsection we derive explicit expressions for the defect contributions to the modified conserved supercharges, momentum and energy.

\subsubsection*{Defect supercharge}

The supersymmetry transformations that leave invariant the bulk action for the $N=1$ sshG model, namely
\br
 \d_s \phi_p &=&\, \vep \psi_p +\bvep \bpsi_p, \label{e3.21}\\
 \d_s \psi_p &=&\, \frac{i\vep}{2} \,(\pa_x-\pa_t)\phi_p + im \bvep \sinh\phi_p,\\
 \d_s \bpsi_p &=& -\frac{i\bvep}{2} \, (\pa_x+\pa_t)\phi_p -im\vep\,\sinh\phi_p,\label{e3.23}
\er
with $p=1,2$, lead to the associated bulk conserved supercharges $Q_\vep$ and $\bq_{\bvep}$, which are given as integrals of local fermionic densities as follows,
\br
 Q_{\vep} &=&\int_{-\infty}^{\infty}dx\,\left(\psi\,(\pa_t-\pa_x)\phi + 2m\bpsi\sinh\phi\),\\
 \bar{Q}_{\bvep}&=& \int_{-\infty}^{\infty}dx\,\left(\bpsi\,(\pa_{t}+\pa_x)\phi- 2m\psi\sinh\phi\). 
\er
However, when the defect is introduced into the theory we have seen that the variation of the bulk action lead to surfaces terms, and then the variation of the defect Lagrangian has to cancel them out exactly  to preserve $N=1$ SUSY. To do that, is necessary to use the following supersymmetry transformations of the degrees of freedom at the defect, 
\br 
 \d_s \l_0 &=& \vep \Big(\psi_++\om_1e^{\l_0\over 2}\cosh\Big(\frac{\ph_-}{2}\Big) f_1\Big)- \bvep\om_2 \,e^{-\frac{(\phi_+-\l_0)}{2}}\cosh\Big(\frac{\ph_-}{2}\Big)\tf_1,\label{e3.24}\\
 \d_s f_1 &=& i\om_1\vep\,e^{\l_0 \over 2}\sinh\Big(\frac{\ph_-}{2}\Big) -\frac{im \bvep}{\om_1}e^{(\ph_+-\l_0) \over 2},\\
 \d_s \tf_1 &=&-\frac{im\vep}{\omega_2} e^{-\frac{\l_0}{2}} -i\om_2 \bvep\, e^{-\frac{(\ph_+-\l_0)}{ 2}}\sinh\Big(\frac{\ph_-}{2}\Big),\label{e3.27}
\er
where $\vep$ and $\bvep$ are Grassmannian parameters. 
Let us compute the associated defect contribution to the supercharges after introducing the defect at $x=0$,
\br
 Q_{\vep} \!&=&\!\int_{-\infty}^{0}\!\!\!dx\left(\psi_1(\pa_t-\pa_x)\phi_1 + 2m\bpsi_1\sinh\phi_1\)\!+\! \int_{0}^{\infty}\!\!\!dx\left(\psi_2(\pa_t-\pa_x)\phi_2 + 2m\bpsi_2\sinh\phi_2\)\!\!,\qquad\,\, \mbox{}\\
 \bq_{\bvep} \!&=&\!\int_{-\infty}^{0}\!\!\!dx\left(\bpsi_1(\pa_{t}+\pa_x)\phi_1 - 2m\psi_1\sinh\phi_1\)\!+\! \int_{0}^{\infty}\!\!\! dx\left(\bpsi_2(\pa_{t}+\pa_x)\phi_2 - 2m\psi_2\sinh\phi_2\)\!.
 \er
Taking the time-derivative lead us to
\br
\frac{dQ_{\vep} }{dt} &=& \Big[\psi_1(\pa_x\phi_1-\pa_t\phi_1) +2m\bpsi_1\sinh\phi_1-\psi_2(\pa_x\phi_2-\pa_t\phi_2)-2m\bpsi_2\sinh\phi_2\Big]_{x=0},\label{e4.4}\qquad \mbox{}\\
\frac{d\bq_{\bvep}}{dt}&=& \Big[\bpsi_1(\pa_x\phi_1+\pa_t\phi_1) +2m\psi_1\sinh\phi_1-\bpsi_2(\pa_x\phi_2+\pa_t\phi_2)-2m\psi_2\sinh\phi_2\Big]_{x=0}.\label{e4.5}
\er
Now, by making use of the defect conditions (\ref{e2.28}) --(\ref{e2.34}) we find after some algebra that the right-hand-side of eqs. (\ref{e4.4}) and (\ref{e4.5})  becomes a total time-derivative, and then the modified conserved supercharges take the following form,
\br 
{\cal Q}=Q_\vep +Q_D, \qquad {\cal \bq}=\bq_{\bvep} +\bq_D,
\er
where the defect contributions are given by
\br
Q_{D} &=& -2\om_1\,e^{\frac{\l_0}{2}}\sinh\Big(\frac{\phi_-}{2}\Big)f_1 +\frac{2m}{\om_2}e^{-\frac{\l_0}{2}} \tf_1,\\
\bq_{D} &=&2 \om_2\,e^{-\frac{(\phi_+-\l_0)}{2}}\sinh\Big(\frac{\phi_-}{2}\Big)\tf_1 +\frac{2m}{\om_1}e^{\frac{(\phi_+ -\l_0)}{2}} f_1.\
\er
%



\subsubsection*{Defect Energy-Momentum}

First of all, let us consider  the canonical energy,
\br
 E &=&\int_{-\infty}^{0} dx\left[ \frac{1}{2}(\pa_{x}\phi_{1})^2 +\frac{1}{2}(\pa_{t}\phi_{1})^2  -i\bpsi_{1}\pa_{x}\bpsi_{1}+i\psi_{1}\pa_{x}\psi_{1} +V_1 +W_1\right] + \non\\
 &&+\int_{0}^{\infty} dx\left[ \frac{1}{2}(\pa_{x}\phi_{2})^2 +\frac{1}{2}(\pa_{t}\phi_{2})^2  -i\bpsi_{2}\pa_{x}\bpsi_{2}+i\psi_{2}\pa_{x}\psi_{2} +V_2 +W_2\right],
\er
with the bulk potentials given in eqs. (\ref{equ2.3}). Then, by computing its time-derivative and using the defect conditions (\ref{e2.28})--(\ref{e2.34}), it can be shown that the modified energy given by ${\cal E} = E +E_D$ is conserved, where  the defect contribution $E_D$ is given by,
\br
 E_D &=& \left[B_0^{(+)}(\phi_+-\l_0,\phi_-)  + B_0^{(-)}(\phi_-,\l_0)\] + \frac{i}{2}\left[\bpsi_+\bpsi_- -\psi_+\psi_-\]\non \\&&\!\!\!\! \!+\left[B_1^{(+)}(\phi_+-\l_0,\phi_-,\bpsi_+,f_1,\tf_1)+B_1^{(-)}(\phi_-,\l_0,\psi_+,f_1,\tf_1)\].\label{e3.9}
\er
The conservation of the modified energy does not impose any constraints on the defect boundary potentials. This fact was expected since the time-translation invariance has not been broken.
 Now, let us consider the bulk contribution to the total canonical momentum
\begin{eqnarray}
 P=\int_{-\infty}^{0}\!\!\! dx\!\left(\pa_{t}\phi_{1}\pa_{x}\phi_{1}-i\bpsi_{1}\pa_{x}\bpsi_{1}-i\psi_{1}\pa_{x}\psi_{1}\right)+\int_{0}^{+\infty}\!\!\! dx\!\left(\pa_{t}\phi_{2}\pa_{x}\phi_{2}-i\bpsi_{2}\pa_{x}\bpsi_{2}-i\psi_{2}\pa_{x}\psi_{2}\right).
\end{eqnarray}
Analogously, by taking its time-derivative and using the bulk field equations (\ref{mov1}), we get the following expression,
\br
 \frac{dP}{dt} &=& \,\,\left[ \frac{1}{2}(\pa_{x}\phi_{1})^2+ \frac{1}{2}(\pa_{t}\phi_{1})^2 -i\bpsi_1\pa_t\bpsi_1 - i\psi_1\pa_t\psi_1 - V_1 -W_1\right]_{x=0}\non\\
 && \!\!\!-\left[ \frac{1}{2}(\pa_{x}\phi_{1})^2+ \frac{1}{2}(\pa_{t}\phi_{2})^2 -i\bpsi_2\pa_t\bpsi_2- i\psi_2\pa_t\psi_2- V_2 -W_2\right]_{x=0},
\er
where the asymptotic contributions at $x=\pm\infty$ are neglected. Now, using intensively the defect conditions (\ref{e2.28})--(\ref{e2.34}), the right-hand side of the above equation reduces to a total time-derivative since  the defect potentials satisfy the two following conditions, 
\br
 2\left({\pa_{\phi_-}B_0^{(+)}} \pa_{\l_0}B_0^{(-)}-\pa_{\l_0}B_0^{(+)}{\pa_{\phi_-} B_0^{(-)}}\) =V_1-V_2,\qquad \qquad \qquad \qquad \mbox{} \label{PB1} \\[0.1cm]
 2\left({\pa_{\phi_-} B_0^{(+)}} \pa_{\l_0} B_1^{(-)}-\pa_{\l_0} B_0^{(+)}\pa_{\phi_-} B_1^{(-)}\)-
 2\left(\pa_{\phi_-} B_0^{(-)}\pa_{\l_0}B_1^{(+)}-\pa_{\l_0} B_0^{(-)}\pa_{\phi_-} B_1^{(+)}\)\qquad\mbox{}\non\\
-i\left(\pa_{f_1} B_1^{(+)} \pa_{f_1}B_1^{(-)}+\pa_{\tf_1} B_1^{(+)}\pa_{\tf_1}B_1^{(-)}\)=W_1-W_2.\qquad \qquad \qquad \quad\,\, \mbox{}\label{PB2}
\er
Then we have that, the modified momentum ${\cal P}=P+P_D$ is conserved, where the defect contribution is given by,
\br
 P_D &=& \left[B_0^{(+)}(\phi_+-\l_0,\phi_-)  - B_0^{(-)}(\phi_-,\l_0)\] + \frac{i}{2}\left[\bpsi_+\bpsi_- +\psi_+\psi_-\]\non \\&&\!\!\!\! \!+\left[B_1^{(+)}(\phi_+-\l_0,\phi_-,\bpsi_+,f_1,\tf_1)-B_1^{(-)}(\phi_-,\l_0,\psi_+,f_1,\tf_1)\].\label{e3.12}
\er
Interestingly, the conditions (\ref{PB1}) and (\ref{PB2}) are Poisson bracket  (PB)  relations whenever the defect potentials $B_k^{(\pm)}$ are regarded as functions of $\l_0$,  $f_1$, and $\tf_1$, and their respective conjugate momenta,
\br
 \Pi_{\l_0}=\frac{\phi_-}{2}, \qquad \Pi_{f_1}=-if_1, \qquad \Pi_{\tf_1}=-i\tf_1.
\er
It is worth pointing out that the defect potentials $B_k^{(\pm)}$ given in eqs. (\ref{e3.15})--(\ref{e3.18}) are particular solutions of the above the PB relations. However, these conditions are valid in general for any model with defect Lagrangian given by (\ref{e2.35}).
It is important to point out that the pure 

\noindent bosonic PB relation (\ref{PB1}) has been previously derived in \cite{Corr4}, where several examples for type-II defects were given. In particular, its  solutions allowed to encompass the Tzitz\'eica-Bullough-Dodd model  within the purely bosonic type-II framework.
To our knowledge, beside the defect potentials for the $N=1$ sshG equation (\ref{e3.15})--(\ref{e3.18}), the only already known solution for this pair of PB relations (\ref{PB1}) and (\ref{PB2}) has been given in \cite{Ale6} for the $N=1$ super-Liouville equation. However, it is natural to believe that exists also other possible solutions like in the bosonic case, and therefore it would be interesting to look for more examples of possible supersymmetric extension of type-II integrable defects.

The results obtained for the defect contribution of the lowest conserved quantities provide a necessary condition for integrability. However, a sufficient condition involves also higher modified conservation laws which can be computed from the defect matrix. The computation of the defect matrix associated to the type-II defect for the sshG model is presented in \mbox{appendix \ref{appB}.}


\subsection{Bosonic limit: Type-II  defect sinh-Gordon model}

Now let us consider the case where the fermionic fields completely vanish. In this case, the bulk Lagrangian densities 
\br
{\cal L}_p =\frac{1}{2}(\pa_x\phi_p)^2-\frac{1}{2}(\pa_t\phi_p)^2 + m^2 (\cosh(2\phi_p)-1), \qquad p=1,2,
\er
describe the sinh-Gordon model, and the defect Lagrangian density can be rewritten as follows
\br
   {\mathcal L}_D =\phi_-\pa_t\l_0 -\frac{1}{2}\phi_-\partial_t\phi_+ + B_0^{(+)}+B_0^{(-)}.\label{ld3.63}
\er 
Here, the defect potentials are
\br
   B_0^{(+)} &=& \om_2^2\, e^{-(\phi_+-\l_0)}\sinh^2\Big(\frac{\phi_-}{2}\Big) +  \frac{m^2}{\om_1^2}e^{(\phi_+-\l_0)}, \label{e3.64} \\
  B_0^{(-)} &=& \om_1^2\, e^{\l_0}\sinh^2\Big(\frac{\phi_-}{2}\Big)+  \frac{m^2}{\om_2^2}e^{-\l_0},\label{e3.65}
\er
and the defect conditions can be rewritten in terms of fields $(\phi_{+},\phi_-,\l_0)$  as follows,
\br
 \pa_x\phi_+ -\pa_t(\phi_+-2\l_0) &=& -\frac{\om_1^2}{2}\,e^{\l_0}\sinh\phi_- -\frac{\om_2^2}{2}\,e^{-(\phi_+-\l_0)}\sinh\phi_-,\\
 \pa_x\phi_- -\pa_t \phi_- &=& -2\om_1^2\, e^{\l_0}\sinh^2\Big(\frac{\phi_-}{2}\Big)+  \frac{2m^2}{\om_2^2}e^{-\l_0},\\
  \pa_x\phi_- +\pa_t \phi_- &=& 2\om_2^2\, e^{-(\phi_+-\l_0)}\sinh^2\Big(\frac{\phi_-}{2}\Big) - \frac{2m^2}{\om_1^2}e^{(\phi_+-\l_0)}.
\er
It is worth pointing out that the dependence on the parameters $(\om_1,\om_2)$ is slightly different from the choice made originally in \cite{Corr4}, where the defect potentials are described by the pair of parameters $(\s,\t)$. However, note that the defect potentials (\ref{e3.64}) and (\ref{e3.65}) belong to a family of equivalent two-parametric solutions of the constraint (\ref{PB1}). This equivalence class 

\noindent can be understand if we perform the following transformation, \br
 B^{(+)} \to B^{(+)} + \rho^{(+)}, \qquad B^{(-)}\to B^{(-)} + \rho^{(-)}.
\er
Then, if $B^{(\pm)}$ are particular solutions for the bosonic PB relation (\ref{PB1}), then the functions $\rho^{(\pm)}$ have to satisfy the following constraint,
\br
\big(\pa_{\phi_-}B^{(+)}\pa_{\l_0}\rho^{(-)}-\pa_{\l_0}\rho^{(+)}\pa_{\phi_-}B^{(-)}\big)=0.
\er
By considering (\ref{e3.64}) and (\ref{e3.65}), the above relation becomes,
\br
\frac{1}{2}\sinh\phi_-\left[\om_2^2 \,e^{-(\phi_+-\l_0)}\pa_{\l_0}\rho^{(-)} -\om_1^2 \,e^{\l_0}\pa_{\l_0}\rho^{(+)}\right] =0.
\er
The possible solutions for $\rho^{(\pm)}$ characterize the equivalence class of the defect potentials $B_0^{(\pm)}$. In particular, if we consider the following functions,
\br
 \rho^{(+)} = \om_2^2\,e^{-(\phi_+-\l_0)}\cosh^2\tau, \qquad \rho^{(-)}= \om_1^2 \,e^{\l_0}\cosh^2\tau,
\er
with the parametrization,
\br
 \om_1^2= \sqrt{2}m\s, \qquad \om_2^2 = \frac{\sqrt{2}m}{\s}, 
\er
we obtain exactly the choice for the defect potentials made in \cite{Corr4}. It is worth pointing out that this equivalence class is strongly reduced in the supersymmetric case, because of the additional PB relation (\ref{PB2}). In that case, we should  perfom also a suitable transformation on the defect potentials $B_1^{(\pm)}$ (\ref{e3.17}) and (\ref{e3.18}), since extra terms coming from derivatives $\rho^{(\pm)}$ would appear and need to be compensated to satisfy (\ref{PB2}). In section \ref{fusing}, it will be seen how the ``compensating'' terms appear naturally by performing a fusing of two kind of defects previously introduced in \cite{FLZ1} for the sshG model.


\subsection{Fermionic limit}
\label{fermi}

A fermionic free field theory is obtained by setting the bosonic fields up to zero directly in the total Lagrangian. Then, for the bulk Lagrangian we get
\begin{eqnarray}
 {\mathcal L}_p &=&  i\psi_p(\partial_x +\partial_t)\psi_p - i\bar{\psi}_p(\partial_x -\partial_t)\bar{\psi}_p -4im \bpsi_p\psi_p,\qquad p=1,2.\mbox{}
\er
The bulk fields equations are given by,
\begin{eqnarray}
(\partial_{x}-\partial_{t})\bar{\psi}_{p}=-2m\,\psi_{p},\qquad
(\partial_{x}+\partial_{t})\psi_{p} = -2m\,\bar{\psi}_{p}.
\end{eqnarray}
The defect Lagrangian takes the following form
\br
   {\mathcal L}_D = \frac{i}{2}(\bar{\psi}_+\bar{\psi}_- -\psi_+\psi_- ) + if_1\pa_t f_1 +i\tf_1\pa_t\tf_1 +B_1^{(+)} +B_1^{(-)},
\er
where
\br
 B_1^{(+)} = \frac{im}{\om_1}\left(\bar{\psi}_+ +{\om_2}\tf_1\right)f_{1}, \qquad
{B}_1^{(-)} =- \frac{im}{\om_2}\big(\psi_+ +{\om_1}f_1\big)\tilde{f}_{1}.\quad\mbox{}\label{fl3.58}
\er
The corresponding defect conditions at $x=0$ take the form,
\br
\psi_- &=& -\frac{m}{\om_2}\tilde{f}_{1}, \qquad \pa_t f_1 = \frac{m}{2\om_1}\bpsi_++\frac{m}{2}\left[\frac{\om_1}{\om_2}+\frac{\om_2}{\om_1}\]\tf_1,\\
\bp_- &=& -\frac{m}{\om_1}f_{1}, \qquad \pp_{t}\tf_{1} = -\frac{{m}}{2\omega_{2}}\psi_+-\frac{m}{2}\left[\frac{\om_1}{\om_2}+\frac{\om_2}{\om_1}\]f_1.
\er
However, note that the defect potentials (\ref{fl3.58}) are not solutions of the PB relation (\ref{PB2}) in the fermionic limit, namely,
\br
-i\left(\pa_{f_1} B_1^{(+)} \pa_{f_1}B_1^{(-)}+\pa_{\tf_1} B_1^{(+)}\pa_{\tf_1}B_1^{(-)}\)=W_1-W_2.
\er
It is not difficult to show that the only possible consistent solution is given for $\tf_1 =\pm f_1$. This case reduces to the fermionic free field model previously discussed in \cite{FLZ1}. After imposing this constrain, the defect Lagrangian becomes,
\br
   {\mathcal L}_D = \frac{i}{2}(\bar{\psi}_+\bar{\psi}_- -\psi_+\psi_- ) + 2if_1\pa_t f_1  + {im}\left[ \om_1^{-1}\bar{\psi}_+ \mp \om_2^{-1}\psi_+\right]f_1 ,\label{fl3.62}
\er
with the following defect conditions
\br
\psi_- &=& \mp\frac{m}{\om_2}{f}_{1}, \qquad
\bp_- = -\frac{m}{\om_1}f_{1},\qquad 
\pa_t f_1 =\frac{m}{4}\left[ \om_1^{-1}\bar{\psi}_+ \mp \om_2^{-1}\psi_+\right].\label{fl3.64}
\er
Note also  that the exact equivalence with the results derived in \cite{FLZ1} would require to perform the transformation  $\psi_2\to-\psi_2$ in expressions (\ref{fl3.62})--(\ref{fl3.64}), together with the following reparametrizations,\vspace{-0.3cm}
\br
 m\to -\frac{m}{2}, \qquad \om_1 \to \frac{\beta\sqrt{m}}{2\sqrt{2}}, \qquad \om_2\to\pm\frac{\sqrt{m}}{\sqrt{2}\beta}.
\er


\section{Fusing defects}
\label{fusing}

In this section  we will construct a fused defect for the $N=1$ sshG model by considering initially a two defect system of the type-I  introduced in \cite{FLZ1, Nathaly1} at different points, and then fusing them to the same point by taking a limit in the  Lagrangian density. Let us consider that one of the defects in placed at $x=0$ and the other at $x=x_0>0$. The Lagrangian density for this system is given by,
\br
 {\mathcal L} = \theta(-x)  {\mathcal L}_1  +\delta(x){\mathcal L}_{D_1} + \theta(x)\th(x_0-x){\cal L}_0 -\delta(x-x_0) {\cal L}_{D_2}+ \theta(x-x_0) {\mathcal L}_2, 
\er
where ${\cal L}_p$, with $p=0,1,2$, are the bulk Lagrange densities given by 
\begin{eqnarray}
 {\mathcal L}_p &=&\frac{1}{2}(\partial_x \phi_p)^2 - \frac{1}{2}(\partial_t \phi_p)^2 +   i\psi_p(\partial_x +\partial_t)\psi_p - i\bar{\psi}_p(\partial_x -\partial_t)\bar{\psi}_p + m^2\left[\cosh(2\phi_p)-1\]\non \\ && \!\! -4im\bpsi_p\psi_p\cosh\phi_p,\qquad\mbox{}
\er
and the two type-I defect Lagrangian densities at $x=0$ ($k=1$), and $x=x_0$ ($k=2$) can be written as
\br
 {\mathcal L}_{D_k} &=&\frac{1}{2}(\phi_0\partial_t\phi_k-\phi_k\partial_t\phi_0) -i\psi_k\psi_0 -i\bar{\psi}_k\bar{\psi}_0 - (-1)^{k} \left(2ig_k\pa_t g_k + B_0^{(k)} +B_1^{(k)}\), \mbox{}
\er

\noindent where $g_k$ are the auxiliary fermionic fields defined in the respective defect positions. The corresponding defect potentials can be written as\footnote{Let us recall that  equivalence with notation used in \cite{FLZ1} requires that $m\to-m/2$, $\psi\to\sqrt{i}\psi$, $\bpsi\to\sqrt{i}\bpsi$, and $g_k\to\sqrt{i}g_k$.}\vspace{-0.2cm}
\begin{eqnarray}
 B_0^{(k)} &=&m\s_k \cosh(\phi_0+\phi_k)+\frac{m}{\s_k}\cosh(\phi_0-\phi_k) , \\
 B_1^{(k)} &=& 2i\sqrt{m}\Big[\sqrt{\s_k}\cosh\Big(\frac{\phi_0+\phi_k}{2}\Big)g_k(\bar{\psi}_0+\bar{\psi}_k) -\frac{(-1)^{k}}{\sqrt{\s_k}}\cosh\Big(\frac{\phi_0-\phi_k}{2}\Big) g_k(\psi_0-\psi_k)\Big],\qquad\mbox{}
\end{eqnarray}
where $\s_k$, with $k=1,2$ are two free parameters associated two each defect.
\noindent Now, it has been claimed that the defects are fused at Lagrangian level after taking the limit $x_0 \to 0$. In that case, we note that there no longer exists bulk Lagrange ${\cal L}_0$ for the fields $\phi_0, \psi_0,$ and $\bpsi_0$, which only contribute to the total defect  Lagrangian at $x=0$ and then becoming  auxiliary fields. The resulting Lagrangian for the fused defect takes the following form,
\br
 {\mathcal L}_{\mbox{}_{||}} = \theta(-x)  {\mathcal L}_1  +\delta(x){\mathcal L}_{D} + \theta(x) {\mathcal L}_2, \label{f4.6}
\er
with the defect Lagrangian ${\cal L}_D={\cal L}_{D_1} -{\cal L}_{D_2}$ at $x=0$ given by
\br
{\cal L}_D = \frac{1}{2}\left(\phi_0\pa_t\phi_- -\phi_- \pa_t \phi_0\) -i \psi_- \psi_0 -i\bpsi_-\bpsi_0 +2ig_1\pa_tg_1 +2i g_2 \pa_t g_2 +B_0 +B_1, \label{f17}
\er
%
The defects potentials $B_0= B_0^{(1)}+B_0^{(2)}$ and $B_1 = B_1^{(1)}+B_1^{(2)}$ can be written now as follows,
\br
 B_0 &=& \frac{m}{2}\left[ e^{\(\frac{\phi_+}{2}+\phi_0\)}\left(\s_1e^{\frac{\phi_-}{2}}+\s_2e^{-\frac{\phi_-}{2}}\) + e^{-\(\frac{\phi_+}{2}+\phi_0\)}\left(\s_1e^{-\frac{\phi_-}{2}}+\s_2e^{\frac{\phi_-}{2}}\)\right.\non\\
 &&\left. \quad + \,e^{\(\frac{\phi_+}{2}-\phi_0\)}\left(\frac{1}{\s_1} e^{\frac{\phi_-}{2}}+\frac{1}{\s_2}e^{-\frac{\phi_-}{2}}\) + e^{-\(\frac{\phi_+}{2}-\phi_0\)}\left(\frac{1}{\s_1}e^{-\frac{\phi_-}{2}}+\frac{1}{\s_2}e^{\frac{\phi_-}{2}}\) \right],\label{f18}\\[0.1cm]
B_1 &=& i\sqrt{m}\,\Big[\,e^{\big(\frac{\phi_+}{4}+\frac{\phi_0}{2}\big)}  \left( \sqrt{\s_1}\,e^{\frac{\phi_-}{4}}\,g_1(\bpsi_0+\bpsi_1) +\sqrt{\s_2}\,e^{-\frac{\phi_-}{4}}\,g_2(\bpsi_0+\bpsi_2)\)   \non \\
&&\qquad +  e^{-\big(\frac{\phi_+}{4}+\frac{\phi_0}{2}\big)}  \left( \sqrt{\s_1}\,e^{-\frac{\phi_-}{4}}\,g_1(\bpsi_0+\bpsi_1) +\sqrt{\s_2}\,e^{\frac{\phi_-}{4}}\,g_2(\bpsi_0+\bpsi_2)\)   \non \\
&&\qquad + e^{\big(\frac{\phi_+}{4}-\frac{\phi_0}{2}\big)}  \left( \frac{1}{\sqrt{\s_1}}\,e^{\frac{\phi_-}{4}}\,g_1(\psi_0-\psi_1) +\frac{1}{\sqrt{\s_2}}\,e^{-\frac{\phi_-}{4}}\,g_2(\psi_2-\psi_0)\)   \non \\
&&\qquad + e^{-\big(\frac{\phi_+}{4}-\frac{\phi_0}{2}\big)}  \left( \frac{1}{\sqrt{\s_1}}\,e^{-\frac{\phi_-}{4}}\,g_1(\psi_0-\psi_1) +\frac{1}{\sqrt{\s_2}}\,e^{\frac{\phi_-}{4}}\,g_2(\psi_2-\psi_0)\) \Big].\label{f19}
\er
First of all, by considering the defect potential (\ref{f18}) of the fused defect, and performing the following identifications,
\br
 \phi_0 = \frac{\l}{2}, \qquad \s_1 = \s e^{-\tau}, \qquad \s_2=\s\,e^{\tau},
\er
we get,
\br
 B_0 &=& \,\,m\s \Big[e^{\frac{(\phi_++\l)}{2}} \cosh\Big(\frac{\phi_-}{2}-\tau \Big)+e^{-\frac{(\phi_++\l)}{2}} \cosh\Big(\frac{\phi_-}{2}+\tau \Big)\Big]\non \\
 &&+ \frac{m}{\s}\Big[e^{\frac{(\phi_+-\l)}{2}} \cosh\Big(\frac{\phi_-}{2}+\tau \Big)+e^{-\frac{(\phi_+-\l)}{2}} \cosh\Big(\frac{\phi_-}{2}-\tau \Big)\Big],
\er

\noindent that the bosonic part of the fused defect described in eqs. (\ref{f17})--(\ref{f19}) is exactly the type-II defect Lagrangian for the sinh-Gordon model introduced in \cite{Corr10}. In this setting, there are no coupling terms for the bulk fields $\phi_1$ and $\phi_2$, which can be understood through the fusing procedure. In order to show the equivalence to the original framework for the type-II defects introduced in \cite{Corr4}, the auxiliary field can be redefined as follows,
\br
 \phi_0 \to -\l_0+\frac{\phi_+}{2}-\ln\left[\cosh\Big(\frac{\phi_-}{2}-\tau\Big) \right].\label{f4.22}
\er
Then the bosonic part of eq. (\ref{f17}) takes the following form,
\br
 {\cal L}_D \Big|_{bosons}\!\! &=& \!\phi_- \pa_t \l_0 -\frac{1}{2}\phi_-\pa_t \phi_+ +
 {m\s}\left[e^{(\phi_+-\l_0)} + e^{-(\phi_+-\l_0)}\cosh\Big(\frac{\phi_-}{2}-\tau\Big)\cosh\Big(\frac{\phi_-}{2}+\tau\Big)\right]\non\\
 && + \frac{m}{\s}\left[e^{-\l_0} + e^{\l_0}\cosh\Big(\frac{\phi_-}{2}-\tau\Big)\cosh\Big(\frac{\phi_-}{2}+\tau\Big)\right],\label{f4.23}
\er
which corresponds to the  bosonic limit (\ref{ld3.63})--(\ref{e3.65}) of the type-II defect Lagrangian density for the sshG model derived through the super-B\"acklund transformation when $\tau =\frac{i\pi}{2}$, and by identifying the parameters as follows, 
\br
 \om_2^{2}=\s^2\om_1^2= m\s.
\er
Let us now consider the fermionic part of the fused defect Lagrangian and discuss  the possibility of obtaining in any limit the type-II defect proposed in section \ref{Lagrangian}. Since the fields $\psi_0$ and $\bpsi_0$ do not have any bulk contribution to the fused Lagrangian ${\cal L}_{\mbox{}_{||}}$ (\ref{f4.6}), and no time-derivatives of them are present in the defect Lagrangian (\ref{f17}), they are essentially Lagrange multipliers which can be eliminated from the defect Lagrangian. To do that, we use the defect equations involving the fields $\psi_0$ and $\bpsi_0$ when $x_0\to 0$, to get
\br
 \psi_0 &=&-\frac{\psi_+}{2} +\sqrt{m}\left[\frac{1}{\sqrt{\s_1}}\cosh\Big(\frac{\phi_0-\phi_1}{2}\Big)g_1+\frac{1}{\sqrt{\s_2}}\cosh\Big(\frac{\phi_0-\phi_2}{2}\Big)g_2\right],\\[0.1cm]
 \bpsi_0 &=& \frac{\bpsi_+}{2} -\sqrt{m}\left[{\sqrt{\s_1}}\cosh\Big(\frac{\phi_0+\phi_1}{2}\Big)g_1\,-{\sqrt{\s_2}}\,\cosh\Big(\frac{\phi_0+\phi_2}{2}\Big)g_2\right].
\er
Now, by noting that
\br
-i\psi_-\psi_0 &=&-\frac{i}{2}\psi_+\psi_- -\frac{2im}{\s}\left[\cosh\Big(\frac{\phi_-}{2}\Big)+\cosh\Big(\frac{\phi_+}{2}-\phi_0\Big)\right]g_1g_2,\\
-i\bpsi_-\psi_0 &=&\,\,\frac{i}{2} \bpsi_+\bpsi_- -{2im\s}\left[\cosh\Big(\frac{\phi_-}{2}\Big)+\cosh\Big(\frac{\phi_+}{2}+\phi_0\Big)\right]g_1g_2,
\er
we find that the fermionic part of the defect Lagrangian  (\ref{f17}) takes the following form,
\br
{\cal L}_D\Big|_{fermions} =  \frac{i}{2}(\bar{\psi}_+\bar{\psi}_- - \psi_+\psi_-) +2ig_1\pa_tg_1 +2i g_2 \pa_t g_2 +B'_1, \label{f4.27}
\er
where, after using the identifications $\s_1=\s e^{-\t}$ and $\s_2=\s e^{\t}$ and performing the shift (\ref{f4.22}) in the field $\phi_0$,  the defect potential $B'_1$ reads,
\br
 B'_1 \! &=& \! i\sqrt{m\s}\Big[ e^{\frac{\left(\phi_+-\l_0\)}{2}}\!\big(\mu_1(\phi_-,\tau)  g_1 + \mu_2(\phi_-,\tau) g_2\big) + e^{-\frac{\left(\phi_+-\l_0\)}{2}}\!\big(\nu_1(\phi_-,\tau)  g_1 + \nu_2(\phi_-,\tau) g_2\big) \Big]\bpsi_+\non \\
 &&\!\!\! \!- \frac{i\sqrt{m}}{\sqrt{\s}}\Big[ e^{\frac{\l_0}{2}}\big(\nu_2(\phi_-,\tau)  g_1 - \nu_1(\phi_-,\tau) g_2\big)+ e^{-\frac{\l_0}{2}}\big(\mu_2(\phi_-,\tau)  g_1 -\mu_1(\phi_-,\tau) g_2\big) \Big]\psi_+ \qquad \qquad \,\mbox{}\non \\
 && \!\!\!\! +2im\Big(\s+\frac{1}{\s}\Big)\cosh\Big(\frac{\phi_-}{2}\Big)g_1g_2 + im\s\Bigg[\frac{e^{(\phi_+-\l_0)}}{\cosh\big(\frac{\phi_-}{2}-\t\big)} + {e^{-(\phi_+-\l_0)}}{\cosh\Big(\frac{\phi_-}{2}-\t\Big)}\Bigg] g_1 g_2\non\\
 && \!\!\!\! +\frac{im}{\s}\Bigg[\frac{e^{-\l_0}}{\cosh\big(\frac{\phi_-}{2}-\t\big)} + {e^{\l_0}}{\cosh\Big(\frac{\phi_-}{2}-\t\Big)}\Bigg] g_1 g_2.
\er
Here we have defined the functions,
\br
 \mu_1 &=&\left[\frac{2}{1+e^{-(\phi_--2\t)}}\right]^{1/2}, \qquad \qquad\,\mu_2 =\left[\frac{2}{1+e^{(\phi_--2\t)}}\right]^{1/2},\label{eq4.32}\\
 \nu_1 &=& \left[\frac{e^{-\phi_-}+e^{-2\t}}{2}\right]^{1/2}={e^{-\t}\over \mu_1}, \qquad \nu_2=\left[\frac{e^{\phi_-}+e^{2\t}}{2}\right]^{1/2}={e^\t \over \mu_2},
\er
satisfying the following relations,
\br
 \mu_1^2+\mu_2^2 &=& 2 , \qquad \,\, \mu_1\nu_2 = e^{\phi_- \over 2}, \qquad \,\, \mu_2\nu_1 = e^{-{\phi_- \over 2}},\label{f4.31}\\
 \qquad \nu_1^2+\nu_2^2 &=& 2\cosh\Big(\frac{\phi_-}{2}-\t\Big)\cosh\Big(\frac{\phi_-}{2}+\t\Big)=\big[\cosh\phi_- +\cosh(2\t)\big].	\label{f4.32}
\er
In order to match to the type-II defect potential $B_1=B_1^{(+)}+B_1^{(-)}$ given by (\ref{e3.17}) and (\ref{e3.18}), the fermionic fields $g_1$ and $g_2$ must be redefined by a linear transformation in terms on new fermionic fields $f_1$ and $\tf_1$. Taking into account the relations (\ref{f4.31}) and (\ref{f4.32}), we propose a general linear transformation as follows,
\br
  g_1 =  \frac{\mu_1}{2} f_1 - \frac{\mu_2}{2}\tf_1, \qquad g_2 =  \frac{\mu_2}{2}f_1 + \frac{\mu_1}{2}\tf_1.
\er
As a result the fermionic part (\ref{f4.27}) of the defect Lagrangian becomes,
\br
{\cal L}_D\Big|_{fermions} \! &=&\!\frac{i}{2}(\bar{\psi}_+\bar{\psi}_- - \psi_+\psi_-)  +if_1\pa_tf_1 +i \tf_1 \pa_t \tf_1 -i\big(\mu_1\pa_t \mu_2 -\mu_2\pa_t\mu_1\big)f_1\tf_1 \non \\
&&\!\! \!-i\sqrt{m\s}\left[\Big(e^{\frac{\left(\phi_+-\l_0\)}{2}}+e^{-\frac{\left(\phi_+-\l_0\)}{2}}\cosh\tau\Big)\bpsi_+f_1 + e^{-\frac{\left(\phi_+-\l_0\)}{2}}\sinh\Big(\frac{\phi_-}{2}\Big)\bpsi_+\tf_1\right]\non\\
&&\!\!\! -i\sqrt{\frac{m}{\s}}\left[\Big(e^{-{\l_0\over 2}}+e^{{\l_0\over 2}}\cosh\t\Big)\psi_+\tf_1-e^{{\l_0\over 2}}\sinh\Big(\frac{\phi_-}{2}\Big)\psi_+f_1\right]\non \\
 && \!\!\!\! +im\Big(\s+\frac{1}{\s}\Big)\!\cosh\Big(\frac{\phi_-}{2}\Big)f_1\tf_1 +\frac{im}{2\s}\Bigg[\frac{e^{-\l_0}}{\cosh\big(\frac{\phi_-}{2}-\t\big)} + {e^{\l_0}}{\cosh\Big(\frac{\phi_-}{2}-\t\Big)}\Bigg]\!f_1\tf_1\non\\
 && \!\!\!\! + {im\s\over 2}\Bigg[\frac{e^{(\phi_+-\l_0)}}{\cosh\big(\frac{\phi_-}{2}-\t\big)} + {e^{-(\phi_+-\l_0)}}{\cosh\Big(\frac{\phi_-}{2}-\t\Big)}\Bigg] f_1\tf_1.\label{fd4.27}
\er
Now, by using the definition of the functions $\mu_{1}$ and $\mu_2$ in (\ref{eq4.32}), we find
\br
 -i\big(\mu_1\pa_t \mu_2 -\mu_2\pa_t\mu_1\big)f_1\tf_1 = \frac{i}{2\cosh\big(\frac{\phi_-}{2}-\t\big)}{(\pa_t\phi_-)}f_1\tf_1.\label{fd4.28}
\er
Then, the above results are suggesting the introduction of a new shift  in the $\l_0$ field as follows,
\br
 \l_0 \to \l_0 + \frac{i}{2\cosh\big(\frac{\phi_-}{2}-\t\big)} f_1\tf_1.\label{f4.37}
\er
Clearly, this redefinition of the auxiliary field $\l_0$ does not contribute with any additional term to the fermionic part of the defect Lagrangian (\ref{fd4.27}). However,  by performing the shift (\ref{f4.37}) over the bosonic part of the defect Lagrangian (\ref{f4.23}), we get the following additional terms,
\br 
 &&-\frac{i}{2\cosh\big(\frac{\phi_-}{2}-\t\big)}{(\pa_t\phi_-)}f_1\tf_1 -
\frac{im\s}{2}\left[\frac{e^{(\phi_+-\l_0)}}{\cosh\big(\frac{\phi_-}{2}-\t\big)} - e^{-(\phi_+-\l_0)}\cosh\Big(\frac{\phi_-}{2}+\tau\Big)\right]f_1\tf_1\non\\
 && - \frac{im}{2\s}\left[\frac{e^{-\l_0}}{{\cosh\big(\frac{\phi_-}{2}-\t\big)}} - e^{\l_0}\cosh\Big(\frac{\phi_-}{2}+\tau\Big)\right]f_1\tf_1.
\er
The first term will cancel out the contribution coming from eq. (\ref{fd4.28}), and the last two terms will contribute to the final form of the defect Lagrangian density, which now can be written in the following way,
\br
 {\cal L}_{D} &=&  \!\phi_- \pa_t \l_0 -\frac{1}{2}\phi_-\pa_t \phi_+ + \frac{i}{2}(\bar{\psi}_+\bar{\psi}_- - \psi_+\psi_-) + if_1\pa_t f_1 +i\tf_1\pa_t\tf_1\nonumber \\[0.1cm]
  &&\,\,+B_0^{(+)} + B_0^{(-)}  + B_1^{(+)}+B_1^{(-)},\label{f4.39}
\er
where the defect potentials are now given by,
\br
 B_0^{(+)}&=&  {m\s}\left[e^{(\phi_+-\l_0)} + e^{-(\phi_+-\l_0)}\Big(\sinh^2\Big(\frac{\phi_-}{2}\Big) +\cosh^2\tau \Big)\right],\label{f4.40}\\
B_0^{(-)} &=& \frac{m}{\s}\left[e^{-\l_0} + e^{\l_0}\Big(\sinh^2\Big(\frac{\phi_-}{2}\Big) +\cosh^2\tau \Big)\right],\\[0.1cm]
 B_1^{(+)} &=& -i\sqrt{m\s}\left[\Big(e^{\frac{\left(\phi_+-\l_0\)}{2}}+e^{-\frac{\left(\phi_+-\l_0\)}{2}}\cosh\tau\Big)\bpsi_+f_1 +e^{-\frac{\left(\phi_+-\l_0\)}{2}}\sinh\Big(\frac{\phi_-}{2}\Big)\bpsi_+\tf_1\right]\non\\
  && + im \s\Big( 1+ {e^{-(\phi_+-\l_0)}} \cosh \tau \Big)\cosh\Big(\frac{\phi_-}{2}\Big)f_1\tf_1, \\[0.1cm]
B_1^{(-)}&= &-i\sqrt{\frac{m}{\s}}\left[\Big(e^{-{\l_0\over 2}}+e^{{\l_0\over 2}}\cosh\t\Big)\psi_+\tf_1-e^{{\l_0\over 2}}\sinh\Big(\frac{\phi_-}{2}\Big)\psi_+f_1\right]\non \\
 && + \frac{ im}{\s}\Big( 1+  {e^{\l_0}}\cosh \tau \Big)\cosh\Big(\frac{\phi_-}{2}\Big)f_1\tf_1.\label{f4.43}
\er
Then, this is the supersymmetric extension of the type-II defect derived in \cite{Corr10} for the sshG model obtained through fusing of two type-I defects of the kind given in \cite{FLZ1}. This defect Lagrangian also contains two arbitrary parameters $\s$ and $\tau$. In addition, it should be pointed out that the type-II defect Lagrangian (\ref{e2.35})--(\ref{e3.18})
derived from consistency with the type-II super-B\"acklund transformation for sshG model can be recovered in the limit $\tau={i\pi/2}$,  after identifying the  parameters
\br
 \om_1 = -\sqrt{\frac{m}{\s}}, \qquad \om_2 =\sqrt{m\s}. \label{f4.44}
\er
Now, the corresponding defect conditions $x=0$ can be written as follows,
\br
 \pa_x \phi_+ -\pa_t(\phi_+ -2\l_0) \!&=&\!  -m\left[\s e^{-(\phi_+-\l_0)}+\frac{1}{\s}e^{\l_0}\]\!\sinh\phi_-  -im\Big(\s+\frac{1}{\s}\Big)\sinh\Big(\frac{\phi_-}{2}\Big) f_1\tf_1\non\\
&&  +i\sqrt{m\s}\, e^{-\frac{\left(\phi_+-\l_0\)}{2}}\cosh\Big(\frac{\phi_-}{2}\Big)\bpsi_+\tf_1 -i\sqrt{m\over \s}e^{\l_0 \over 2}\cosh\Big(\frac{\phi_-}{2}\Big)\psi_+f_1\qquad\mbox{} \non\\
&&\!-im\left[\s e^{-(\phi_+-\l_0)} +\frac{1}{\s}e^{\l_0}\right]\!\cosh\t\sinh\Big(\frac{\phi_-}{2}\Big) f_1\tf_1,\qquad \mbox{}\label{f4.45}
\er
 \br
 (\pa_x+\pa_t)\phi_- &=& 2m\s\left[ e^{-(\phi_+-\l_0)}\Big(\sinh^2\Big(\frac{\phi_-}{2}\Big) +\cosh^2\tau \Big)-e^{(\phi_+-\l_0)} \right]\non\\
 && +i\sqrt{m\s}\Big(e^{\frac{\left(\phi_+-\l_0\)}{2}}-e^{-\frac{\left(\phi_+-\l_0\)}{2}}\cosh\tau\Big)\bpsi_+f_1 \non \\
 && -i\sqrt{m\s}\,e^{-\frac{\left(\phi_+-\l_0\)}{2}}\sinh\Big(\frac{\phi_-}{2}\Big)\bpsi_+\tf_1\non\\
 && +2im\s \,e^{-(\phi_+-\l_0)}\cosh\t \cosh\Big(\frac{\phi_-}{2}\Big)f_1\tf_1,\\[0.1cm]
 (\pa_x-\pa_t)\phi_-   &=& \frac{2m}{\s}\left[e^{-\l_0} - e^{\l_0}\Big(\sinh^2\Big(\frac{\phi_-}{2}\Big) +\cosh^2\tau \Big)\right]\non\\
 && -i\sqrt{\frac{m}{\s}}\left[\Big(e^{-{\l_0\over 2}}-e^{{\l_0\over 2}}\cosh\t\Big)\psi_+\tf_1 +e^{{\l_0\over 2}}\sinh\Big(\frac{\phi_-}{2}\Big)\psi_+f_1\right]\non\\
 && -\frac{2im}{\s }\,e^{\l_0}\cosh\t \cosh\Big(\frac{\phi_-}{2}\Big)f_1\tf_1,\\[0.1cm]
 \psi_-  &=& \sqrt{\frac{m}{\s}}\left[e^{{\l_0\over 2}}\sinh\Big(\frac{\phi_-}{2}\Big)f_1-\Big(e^{-{\l_0\over 2}}+e^{{\l_0\over 2}}\cosh\t\Big)\tf_1\right],\\[0.1cm]
 \bpsi_- &=&\sqrt{m\s}\!\left[\!\Big(e^{\frac{\left(\phi_+-\l_0\)}{2}}\!+e^{-\frac{\left(\phi_+-\l_0\)}{2}}\cosh\tau\Big)f_1 +e^{-\frac{\left(\phi_+-\l_0\)}{2}}\sinh\Big(\frac{\phi_-}{2}\Big)\!\tf_1\right]\!\!,\quad \mbox{}\\[0.1cm]
 \pa_t f_1 &=& -\frac{\sqrt{m\s}}{2}\Big(e^{\frac{\left(\phi_+-\l_0\)}{2}}+e^{-\frac{\left(\phi_+-\l_0\)}{2}}\cosh\tau\Big)\bpsi_+ +\frac{1}{2}\sqrt{\frac{m}{\s}}
 e^{{\l_0\over 2}}\sinh\Big(\frac{\phi_-}{2}\Big)\psi_+\non \\
 &&-\frac{m}{2} \left[\Big( \s+\frac{1}{\s}\Big) + \Big(\s {e^{-(\phi_+-\l_0)}} +\frac{1}{\s}  {e^{\l_0}}\Big)\cosh \tau\right]\!\cosh\Big(\frac{\phi_-}{2}\Big)\tf_1, \\[0.1cm]
 \pa_t\tf_1   &=&  -\frac{\sqrt{m\s}}{2}e^{-\frac{\left(\phi_+-\l_0\)}{2}}\sinh\Big(\frac{\phi_-}{2}\Big)\bpsi_+ -\frac{1}{2}\sqrt{\frac{m}{\s}}\Big(e^{-{\l_0\over 2}}+e^{{\l_0\over 2}}\cosh\t\Big)\psi_+\non \\
 &&  +\frac{m}{2}  \left[\Big( \s+\frac{1}{\s}\Big) + \Big(\s {e^{-(\phi_+-\l_0)}} +\frac{1}{\s}  {e^{\l_0}}\Big)\cosh \tau\right]\!\cosh\Big(\frac{\phi_-}{2}\Big)f_1.\label{f4.51}
\er
It is easy to see that defect conditions (\ref{e2.28})--(\ref{e2.34}) are recovered in the limit $\tau=i\pi/2$ using the identifications (\ref{f4.44}). In addition,  it can be shown directly that  the type-II defect potentials obtained by fusing procedure (\ref{f4.40})--(\ref{f4.43}) also satisfy the Poisson bracket constraints (\ref{PB1}) and (\ref{PB2}), guaranteeing that they belong to the two-parametric family of equivalent solutions, and consequently the modified energy, momentum and supersymmetry of the non-fused type-I defects are preserved after fusing them. This  fact  indicates classical integrability of the fused defect. Nevertheless, it remains to derive the corresponding defect matrix associated to the defect conditions obtained by fusing procedure (\ref{f4.45})--(\ref{f4.51}).  It allows us to compute defect contributions to an infinite set of modified conserved quantities for the sshG model. The explicit computations are presented in appendix \ref{appB}.

%
\section{B\"acklund solution for the sshG model}
\label{solutions}
In this section, we will consider the classical behaviour of a solution of the sshG equation in the presence of the type-II defect.

Let us first consider the following solution for the $N=1$ sshG equation on each side of the defect \cite{FLZ1},
\be\label{solsuper}
e^{\ph_1}&=&\frac{1+E_1}{1-E_1},\qquad e^{\ph_2}=-\frac{1+E_2}{1-E_2},\qquad E_j=R_{j}\, e^{ax+bt},\quad j=1,2\non\\
\bp_j&=& \epsilon\, s_j\, e^{ax+bt} \left(\frac{1}{1+E_j}+\frac{1}{1-E_j}\right),\quad \p_1=e^{\theta}\bp_1,\quad \p_2=-e^{\theta}\bp_2.
\ee
where $\epsilon$ carries the Grassmannian character of the Majorana fields, $z=R_2/R_1$ represents the delay suffered by the bosonic part of the solution, as well as $\zeta=s_2/s_1$ represents the delay suffered by the fermionic part of it. The constant $a$ and $b$ satisfy the relation $a^2-b^2=4m^2$, and then can be conveniently chosen as,
\be
a=-2m\cosh\theta, \quad b=2m\sinh\theta,
\ee
where $\theta>0$ is the rapidity parameter. Note also that the solution (\ref{solsuper}) only  have one fermionic parameter $\eps$, and therefore bilinear fermionic terms immediately  vanish. In fact, the fermionic solutions are not involved in the computation of the delay $z$ obtained from  the bosonic defect conditions,
\be
(\pp_{x}+\pp_{t})\ph_- &=& 2\o_2^{2}e^{-(\ph_{+}-\l_0)}\sinh^2\Big(\frac{\ph_-}{2}\Big)-\frac{2m^2}{\o_1^{2}}e^{(\ph_{+}-\l_0)}, \label{bs5.23}\\
(\pp_{x}-\pp_{t})\ph_- &=& -2\o_1^{2}e^{\l_0}\sinh^2\Big(\frac{\ph_-}{2}\Big)+\frac{2m^2}{\o_2^{2}}e^{-\l_0}.\label{bs5.24}
\ee
From eqs. (\ref{bs5.23}) and (\ref{bs5.24}) we get the two possible expressions for the auxiliary field $\l_0$,
\be
e^{-\l_0}&=&-\frac{\o_{2}^{2}\,e^{-\ph_+}(\pp_x-\pp_t)\ph_-+ \o_{1}^{2}(\pp_x+\pp_t)\ph_-}{4m^2\sinh\ph_+},\label{el}\\
e^{\l_0}&=&-\frac{\o_{2}^{2}\,e^{\ph_+}(\pp_x-\pp_t)\ph_-+\o_{1}^{2}(\pp_x+\pp_t)\ph_-}{2\o_{1}^{2}\o_{2}^{2}\sinh\ph_+(\cosh\ph_{-}-1)}.\label{eml}
\ee
Then, we get two possibilities for  the bosonic delay $z$, namely
\be
z_{1}=\left(\frac{e^{\eta _2+\theta }+i e^{\eta _1}
}{e^{\eta_2+\theta }-i e^{\eta_1}}\right)^2, \qquad z_2=  \frac{1}{z_1},
\ee
where we have introduced the parametrization $\om_k=e^{\eta_k}$, $k=1,2$. This expression can be rewritten as 
\be\label{z}
z_1=\tanh\left(\frac{\th+\eta_2-\eta_1}{2}+\frac{i\pi}{4}\right)\coth\left(\frac{\th+\eta_2-\eta_1}{2}-\frac{i\pi}{4}\right).
\ee
Now, by considering $z=z_1$ we find
\be\label{l}
e^{\l_0}&=&\frac{im}{\o_1\o_2}\frac{(1+R_1d) (1+R_2d)}{(1-\rho R_1d) (1-\tilde{\rho} R_1d)},
\ee
where,
\be
\rho=\tanh\left(\frac{\th+\eta_2-\eta_1}{2}+\frac{i\pi}{4}\right),\qquad\tilde{\rho}=\coth\left(\frac{\th+\eta_2-\eta_1}{2}-\frac{i\pi}{4}\right), \qquad d=e^{bt}.\qquad \mbox{}
\ee
It is important to note that by making $\eta_2 -\eta_1=\eta$, and $\th\to-\th$, we recover results previously analysed in \cite{Corr4, Corr10}.


On the other hand, we can find expressions for the auxiliary fermionic fields in the same way implemented for the field $\l_0$ . In this case, using the equations (\ref{e2.14}) and (\ref{e2.15}) we find the following expressions for the auxiliary fermionic fields $f_1$ and $\tilde{f}_1$,
\be
f_1&=&-\,\left[\frac{m \,e^{-\eta_2}\,\bp_- +e^{-\left(\frac{\phi_+}{ 2}-\eta_2\)}\,e^{\l_0}\sinh\Big(\frac{\ph_-}{2}\Big)\p_-}{ m^2 e^{\frac{(\ph_+-\l_0)}{2}-(\eta_1+\eta_2)}+ e^{-\frac{(\ph_+-\l_0)}{2}+(\eta_1+\eta_2)}\,e^{\l_0}\sinh^2\Big(\frac{\ph_-}{2}\Big)}\],\label{f1}\\
\tilde{f}_1&=&-\,\left[\frac{m \,e^{\left(\frac{\ph_+}{2}-\eta_1\)}\,\psi_- -e^{\eta_1}\,e^{\l_0}\sinh\Big(\frac{\ph_-}{2}\Big)\bp_-}{ m^2 e^{\frac{(\ph_+-\l_0)}{2}-(\eta_1+\eta_2)}+ e^{-\frac{(\ph_+-\l_0)}{2}+(\eta_1+\eta_2)}\,e^{\l_0}\sinh^2\Big(\frac{\ph_-}{2}\Big)}\].\label{tildef1}
\ee
An intriguing result is that the fermionic delay $\zeta$ turns to be  arbitrary, in the sense that the defect conditions do not provide any constraint that allows us to determine $s_2$ in terms of $s_1$, at least for one-soliton solution. The consistency of the defect conditions for the fermionic fields suggests that somehow the auxiliary fields $f_1$ and $\tf_1$ compensate the delay suffered by the fermion fields passing through the  defect. This curious fact could be better understand if one consider more general soliton-solutions for the sshG-model.

However, as it might been expected  from the results obtained in \cite{FLZ1} for a fermion/boson system passing through a  defect, the fermionic delay $\zeta$ would be the same as the bosonic delay $z$. In  particular we would have an additional constraint on the auxiliary fermionic fields. By assuming that  the fermionic delay is given by  
\be
\zeta =z= \left(\frac{e^{\eta _2+\theta }+i e^{\eta _1}
}{e^{\eta_2+\theta }-i e^{\eta_1}}\right)^2 ,
\ee
we find that the two auxiliary fermionic fields $f_1$ and $\tilde{f}_{1}$ are related as,
\be
f_1&=& -\left[{(1+R_1d)(1+zR_1d) \over (1-R_1d)(1-zR_1d) }\right]^{\frac{1}{2}}\left(\frac{1+\sqrt{z}\ R_1d}{1-\sqrt{z}\ R_1d}\right)\tilde{f}_1.\label{5.33}
\ee 
Although $\sqrt{z}$ has two roots, only the negative root is considered for consistency with the fermionic defect conditions in order to obtain the correct result for the fermionic delay. In the fermionic limit ($R_1\to 0$) the auxiliary field $f_1$ becomes $-\tf_1$, which is consistent with results obtained in subsection \ref{fermi}. 

Considerations of a general class of solutions of the type-II defect conditions following the line of \cite{FLZ1}, which could allow changes in the character of the soliton  could give a light to understand the apparent arbitrariness in determining the fermionic delay. This relevant issue raised from this work is important to address future investigations and deserves to be deeply studied.


\section{Final remarks}
\label{final}

In this paper, we have derived a type-II integrable defect for the $N=1$ supersymmetric sinh-Gordon model by using two different methods, namely, superextension of the B\"acklund transformation and the fusing procedure. A new  super-B\"acklund for the $N=1$ sshG equation was given in section \ref{backlund}, described by the three superfields  $\L, f$ and $\tf$. The defect conditions consistent with this new transformation  were derived in section \ref{Lagrangian}, the supersymmetric invariance was properly shown, as well as the derivation of the modified conserved energy and momentum. Interestingly, the bosonic limit corresponds to the type-II B\"acklund transformation for the sinh-Gordon model \cite{Corr4, Ale3}. Also, it has been shown that the type-II defect derived for the sshG model can be obtained by fusing two defects of the kind previously derived in \cite{FLZ1}. 

In view of the results, it would be interesting to consider the possibility of finding supersymmetric extensions of the \mbox{type-II } defects for other models with extended supersymmetry, for instance the $N=2$ super-Liouville and sshG equations. Associated integrable defects could be found as a results of the fusing  defects of the kind already known \cite{FLZ2}. 

Also,  it is possible to find new integrable boundary conditions for the sshG model by using the type-II integrable defect for the sshG following the line of \cite{Zambon1}--\cite{Zambon2}. Some of these questions are expected to be developed in future investigations.


\vskip 1cm
 \noindent
{\bf Acknowledgements} \\
\vskip .1cm
 \noindent ARA would like to thank FAPESP S\~ao Paulo Research Foundation for financial support under the PD Fellowship 2012/13866-3. NIS, JFG and AHZ  thank also CNPq for financial support. We are also grateful to the referee for the careful review, and  helpful  comments and suggestions.

\appendix

\section{Defect matrix}
\label{appB}
Let us consider the Lax pair  ${\cal A}_{\pm}^{(p)}$ depending on the respective fields $\phi_p, \psi_p$, and $\bpsi_p$, written in the following form\footnote{Note that equivalence with notation originally used  in \cite{Nathaly1} would require that $m=-2$, $\psi_p=\sqrt{i}\psi_p$ and $\bpsi_p=\sqrt{i}\bpsi_p$.},
\br
\mathcal{A}_{+}^{(p)} &=&\left(\begin{array}{cc|c}\lambda^{1/2}-\pp_{+}\phi_p&-1&\sqrt{i}\bp_p\\[0.2cm] -\lambda&\lambda^{1/2}+\pp_{+}\phi_p&\l^{1/2}\sqrt{i}\bp_p \tabularnewline\hline \mbox{}&\mbox{}&\mbox{}\\[-0.3cm]
	\l^{1/2}\sqrt{i}\bp_p&\sqrt{i}\bp_p& 2\lambda^{1/2}       \end{array}\right)\label{lax +}, \\[0.3cm]
\mathcal{A}_{-}^{(p)}&=&\left(\begin{array}{cc|c}\frac{m^2}{4}\lambda^{-1/2}& -\frac{m^2}{4 }\l^{-1} e^{2 \phi_p}& \frac{ \sqrt{i} m}{2 } \l^{-1/2}\psi_p \,e^{\phi_p}\\[0.2cm]-\frac{m^2 }{4} e^{-2 \phi_p} & \frac{m^2}{4 } \l^{-1/2}& \frac{\sqrt{i} m}{2} \psi_p  \,e^{-\phi_p}  \tabularnewline\hline \mbox{}&\mbox{}&\mbox{}\\[-0.3cm]
	-\frac{\sqrt{i} m}{2}  \psi_p \,e^{-\phi_p}  & -\frac{ \sqrt{i} m }{2 } \l^{-1/2}\psi_p\,e^{\phi_p}& \frac{m^2}{2 }\l^{-1/2}\end{array}\right)\label{lax -},
\er
where $\l$ here represents the spectral parameter and should not be confused with the any component of the superfield $\L$. The graded matrix $\cK$ connecting two different solutions of the associated linear problem for the $N=1$ sshG equation, namely $\Psi_1=\cK(\l)\Psi_2$, where 
\br
\pa_\pm \Psi_p (x_\pm,\l) =-{\cal A}_\pm^{(p)}\Psi_p(x_\pm,\l), \qquad p=1,2,
\er
satisfies the following equations,
\br\label{gauge}
\pa_{\pm}\cK=\cK\mathcal{A}^{(1)}_{\pm}-\mathcal{A}^{(2)}_{\pm}\cK. 
\er
First of all, to make the derivation of the defect matrices clearer for the reader we will now present the  system of  equations (\ref{gauge}) explicitly in components (see also \cite{Nathaly1}). 
Then, by considering the following  $\l$-expansion for $\cK$,
\begin{eqnarray}\label{K}
\cK_{ij}=\a_{ij}+\l^{-1/2}\b_{ij}+\l^{1/2}\g_{ij},
\end{eqnarray}
where $\a_{ij}, \b_{ij}$, and $\g_{ij}$ being the entries of $3 \times 3$ graded matrices, we get the following:\\\\
{\bf $\lambda^{+3/2}$- terms:}
\begin{eqnarray}
\g_{12}=\g_{13}=\g_{32}=0, \qquad \quad \g_{11}=\g_{22}. \label{dm5}
\end{eqnarray}
{\bf $\l^{+1}$- terms:} 
\begin{eqnarray}
\a_{12} &=& \g_{13}\,\sqrt{i}\bp_1, \label{dm6}\\
\g_{13} &=& -\g_{12}\,\sqrt{i}\bp_1,\label{dm7}\\
\g_{32} &=&-\g_{12}\,\sqrt{i}\bp_2,\label{dm8}\\
\a_{11}-\a_{22}&=&\sqrt{i}\left(\bp_2\,\g_{31}-\g_{23}\,\bp_1\),\label{dm9}\\
\a_{13}+\g_{23}&=&\sqrt{i}\left(\bp_2\,\g_{33}-\g_{11}\,\bp_1\),\label{dm10}\\ 
\g_{31}+\a_{32}&=&\sqrt{i}\left(\bp_1\,\g_{33}-\g_{11}\,\bp_2\).\label{dm11}
\end{eqnarray}
{\bf $\l^{-3/2}$- terms:} 
\begin{eqnarray}
\b_{21}=\b_{23}= \b_{31}=0, \qquad \quad \b_{22}=\b_{11}e^{2\phi_-}.\label{dm12}
\end{eqnarray}
{\bf$\l^{-1}$- terms:} 
\begin{eqnarray}
\a_{21} &=&\frac{2\sqrt{i}}{m}\psi_2\,\b_{31}\,e^{-\frac{(\phi_+-\phi_-)}{2}},\label{dm13}\\
\b_{23} &=&-\frac{2\sqrt{i}}{m}\b_{21}\,\psi_1\,e^{\frac{(\phi_+ +\phi_-)}{2}},\label{dm14}\\ 
\b_{31} &=&\frac{2\sqrt{i}}{m}\psi_2\,\b_{21}\,e^{\frac{(\phi_+-\phi_-)}{2}},\label{dm15}\\
\a_{11}\,e^{\phi_-}-\a_{22}\,e^{-\phi_-}&=&\frac{2\sqrt{i}}{m}e^{\frac{\phi_+}{2}}\big(\a_{23}\,\p_1\,e^{-\frac{\phi_-}{2}}+e^{\frac{\phi_-}{2}}\,\p_2\,\a_{31}\big),	\label{dm16}\\
\a_{11}\,e^{\phi_-}-\a_{22}\,e^{-\phi_-}&=&-\frac{2\sqrt{i}}{m}e^{-\frac{\phi_+}{2}}\big(\p_2\b_{32}\,e^{-\frac{\phi_-}{2}}+e^{\frac{\phi_-}{2}}\b_{13}\,\p_1\big),\label{dm17}\\
\a_{31}\,e^{{(\phi_+ +\phi_-)}}+\b_{32}&=& -\frac{2\sqrt{i}}{m}e^{\frac{\phi_+ }{2}}\big(\b_{33}\,\p_1\,e^{\frac{\phi_-}{2}}-\b_{22}\,\p_2\,e^{-\frac{\phi_-}{2}}\big),\label{dm18}\\
\a_{23}\,e^{(\phi_+-\phi_-)}+ \b_{13}&=&-\frac{2\sqrt{i}}{m}\,e^{\frac{\phi_+}{2}}(\b_{11}\,\p_1e^{\frac{\phi_-}{2}}-\b_{33}\,\p_2 e^{-\frac{\phi_-}{2}}).\label{dm19}
\end{eqnarray}
We can see that eq. (\ref{dm5}) and (\ref{dm6}) implies immediately that $\a_{12}=0$, and similarly from eqs. (\ref{dm12}) and (\ref{dm13}) we find that $\a_{21}=0$. A set of differential equations arise from the $\l^0$ and $\l^{\pm1/2}$ terms, which are presented below:\\

\noindent{\bf$\l^{0}$- terms:} 
\begin{eqnarray}
\pp_{+}\a_{11}&=&-\a_{11}\,\pp_{+}\phi_- +\sqrt{i}(\b_{13}\,\bp_1-\bp_2\,\a_{31}),\label{dm20}\\
\pp_{+}\a_{13}&=&\frac{\a_{13}}{2}\,\pp_{+}({\phi_+-\phi_-})+\b_{13}+\a_{23}-\sqrt{i}\bp_2\,\a_{33}+\sqrt{i} (\a_{11}+\b_{12})\bp_1,\label{dm21}\\
\pp_{+}\a_{23}&=&-\frac{\a_{23}}{2}\pp_{+}({\phi_+-\phi_-})+\sqrt{i}(\b_{22}\,\bp_1-\bp_2\,\b_{33}),\label{dm22}\\
\pp_{+}\a_{22}&=&\a_{22}\,\pp_{+}\phi_- +\sqrt{i}(\a_{23}\,\bp_1-\bp_2\,\b_{32}),\label{dm23}\\
\pp_{+}\a_{31}&=&-\frac{\a_{31}}{2}\pp_{+}({\phi_++\phi_-})+\sqrt{i}(\b_{33}\,\bp_1-\bp_2\,\b_{11}),\label{dm24}\\
\pp_{+}\a_{32}&=&\frac{\a_{32}}{2}\pp_{+}({\phi_++\phi_-})-\a_{31}-\b_{32}+\a_{33}\,\sqrt{i}\bp_1-\sqrt{i}\bp_2(\b_{12}+\a_{22}),\label{dm25}\\
\pp_{+}\a_{33}&=&\sqrt{i}(\b_{32}+\a_{31})\bp_1-\sqrt{i}\bp_2(\a_{23}+\b_{13}),\label{dm26}\\[0.1cm]
\pp_{-}\a_{11}&=&-\frac{\sqrt{i}m}{2}\,e^{-\frac{\phi_-}{2}}(\a_{13}\,\p_1e^{-\frac{\phi_+}{2}}+e^{\frac{\phi_+}{2}}\p_2\,\g_{31}),\label{dm27}\\
\pp_{-}\a_{13}&=&-\frac{\sqrt{i}m}{2}\,e^{\frac{\phi_+}{2}}(e^{-\frac{\phi_-}{2}}\,\p_2\,\g_{33}-\g_{11}\,\p_1\,e^{\frac{\phi_-}{2}}), \label{dm28}\\
\pp_{-}\a_{22}&=&-\frac{\sqrt{i}m}{2}\,e^{\frac{\phi_-}{2}}(e^{-\frac{\phi_+}{2}}\,\p_2\,\a_{32}+\g_{23}\,\p_1\,e^{\frac{\phi_+}{2}}),\label{dm29}
\er
\br
\pp_{-}\a_{23}&=&\frac{m^2}{4}\left(\g_{23}+\a_{13}\,e^{-(\phi_+-\phi_-)}\right)+\frac{\sqrt{i}m}{2}(\g_{21}\,e^{\frac{(\phi_++\phi_-)}{2}}+\a_{22}\,e^{-\frac{(\phi_++\phi_-)}{2}})\,\p_1\non\\&&-\frac{\sqrt{i}m}{2}\,e^{-\frac{(\phi_+-\phi_-)}{2}}\p_2\,\a_{33},\qquad \,\,\,\mbox{}\label{dm30}\\
\pp_{-}\a_{31}&=&-\frac{m^2}{4}\left(\g_{31}+\a_{32}e^{-(\phi_++\phi_-)}\right)+\frac{\sqrt{i}m}{2}\p_2(e^{-\frac{(\phi_+-\phi_-)}{2}}\a_{11}+e^{\frac{(\phi_+-\phi_-)}{2}}\g_{21})\non\\&&-\frac{\sqrt{i}m}{2}\a_{33}\p_1\,e^{-\frac{(\phi_++\phi_-)}{2}},\qquad\mbox{}\label{dm31}\\
\pp_{-}\a_{32}&=&-\frac{\sqrt{i}m}{2}e^{\frac{\phi_+}{2}}(\g_{33}\,\p_1\,e^{\frac{\phi_-}{2}}-e^{-\frac{\phi_-}{2}}\p_2\,\g_{11}),\label{dm32}\\
\pp_{-}\a_{33}&=&\frac{\sqrt{i}m}{2}(\g_{31}e^{\frac{(\phi_++\phi_-)}{2}}+\a_{32}\,e^{-\frac{(\phi_++\phi_-)}{2}})\p_1+\frac{\sqrt{i}m}{2}\p_2(e^{-\frac{(\phi_+-\phi_-)}{2}}\a_{13}+e^{\frac{(\phi_+-\phi_-)}{2}}\g_{23})\label{dm33}.\non\\
\end{eqnarray}
{\bf$\l^{-1/2}$- terms:} 
\begin{eqnarray}
\pp_{-}\b_{11}&=&\frac{m^2}{4}\,e^{-\phi_-}(\g_{21}\,e^{\phi_+}-\b_{12}\,e^{-\phi_+})-\frac{\sqrt{i}m}{2}\,e^{-\frac{\phi_-}{2}}(\b_{13}\,\p_1\,e^{-\frac{\phi_+}{2}}+e^{\frac{\phi_+}{2}}\,\p_2\,\a_{31}),\qquad\mbox{} \label{dm34}\\
\pp_{-}\b_{12}&=&\frac{m^{2}\g_{11}}{4}\,e^{\phi_+}(e^{-\phi_-}-e^{\phi_-})-\frac{\sqrt{i}m}{2}e^{\frac{\phi_+}{2}}(\a_{13\,}\p_1\,e^{\frac{\phi_-}{2}}+e^{-\frac{\phi_-}{2}}\,\p_2\,\a_{32}),\label{dm35}\\
\pp_{-}\b_{13}&=&\frac{m^2}{4}\left(\g_{23}\,e^{(\phi_+-\phi_-)}+\a_{13}\right)+\frac{\sqrt{i}m}{2}(\a_{11}\,e^{\frac{(\phi_++\phi_-)}{2}}+\b_{12}\,e^{-\frac{(\phi_++\phi_-)}{2}})\,\p_1\non\\&&-\frac{\sqrt{i}m}{2}e^{\frac{(\phi_+-\phi_-)}{2}}\p_2\,\a_{33},\qquad \mbox{}\label{dm36}\\
\pp_{-}\b_{22}&=&\frac{m^2}{4}\,e^{\phi_-}(\b_{12}\,e^{-\phi_+}-\g_{21}\,e^{\phi_+})-\frac{\sqrt{i}m}{2}e^{\frac{\phi_-}{2}}(\a_{23}\,\p_1\,e^{\frac{\phi_+}{2}}+e^{-\frac{\phi_+}{2}}\,\p_2\,\b_{32}),\label{dm37}\\	
\pp_{-}\b_{32}&=&-\frac{m^2}{4}\left(\a_{32}+\g_{31}\,e^{(\phi_++\phi_-)}\right)+\frac{\sqrt{i}m}{2}\p_2(e^{\frac{(\phi_+-\phi_-)}{2}}\a_{22}+e^{-\frac{(\phi_+-\phi_-)}{2}}\b_{12})\non\\&&-\frac{\sqrt{i}m}{2}\a_{33}\,\p_1\,e^{\frac{(\phi_++\phi_-)}{2}},\qquad \mbox{}\label{dm38}\\
\pp_{-}\b_{33}&=& \frac{\sqrt{i}m}{2}(\b_{32}e^{-\frac{(\phi_++\phi_-)}{2}}\!+\a_{31}e^{\frac{(\phi_++\phi_-)}{2}})\p_1+\!\frac{\sqrt{i}m}{2}\p_2(e^{-\frac{(\phi_+-\phi_-)}{2}}\b_{13}+e^{\frac{(\phi_+-\phi_-)}{2}}\a_{23}),\non\\ \mbox{}\\
\pp_{+}\b_{11}&=&-\b_{11}\,\pp_{+}\phi_-,\label{dm40}\\
\pp_{+}\b_{12}&=&\b_{12}\,\pp_{+}\phi_+ +\b_{22}-\b_{11}+\sqrt{i}(\b_{13}\,\bp_1-\bp_2\,\b_{32}),\label{dm41}\\
\pp_{+}\b_{13}&=&\frac{\b_{13}}{2}\,\pp_{+}({\phi_+-\phi_-})+\sqrt{i}(\b_{11}\bp_1-\bp_2\b_{33}),\label{dm42}\\%
\pp_{+}\b_{22}&=&\b_{22}\,\pp_{+}\phi_-, \label{dm43}\\[0.1cm]
\pp_{+}\b_{32}&=&\frac{\b_{32}}{2}\pp_{+}({\phi_++\phi_-})+\sqrt{i}(\b_{33}\,\bp_1-\bp_2\,\b_{22}),\label{dm44}\\
\pp_{+}\b_{33}&=&0.\label{dm45}
\end{eqnarray}
{\bf$\l^{+1/2}$- terms :} 
\begin{eqnarray}
\pp_{-}\g_{11}&=& 0, \label{dm46}\qquad\\
\pp_{-}\g_{21}&=&\frac{m^2\g_{11}}{4}\,e^{-\phi_+}(e^{\phi_-}-e^{-\phi_-})-\frac{\sqrt{i}m}{2}e^{-\frac{\phi_+}{2}}(\g_{23}\,\p_1\,e^{-\frac{\phi_-}{2}}+e^{\frac{\phi_-}{2}}\,\p_2\,\g_{31}),\label{dm47}\\
\pp_{-}\g_{23}&=&-\frac{\sqrt{i}m}{2}e^{-\frac{\phi_+}{2}}(e^{\frac{\phi_-}{2}}\,\p_2\,\g_{33}-\g_{11}\,\p_1\,e^{-\frac{\phi_-}{2}}),\label{dm48}\\
\pp_{-}\g_{31}&=&-\frac{\sqrt{i}m}{2}e^{-\frac{\phi_+}{2}}(\g_{33}\,\p_1\,e^{-\frac{\phi_-}{2}}-e^{\frac{\phi_-}{2}}\p_2\,\g_{11}), \label{dm49}\\
\pp_{-}\g_{33}&=& 0,\qquad \label{dm50}\\
\pp_{+}\g_{11}&=&-\g_{11}\,\pp_{+}\phi_- +\g_{21}-\b_{12}+\sqrt{i}(\a_{13}\,\bp_1-\bp_2\,\g_{31}),\label{dm51}\\
\pp_{+}\g_{22}&=&\g_{22}\,\pp_{+}\phi_- -\g_{21}+\b_{12}+\sqrt{i}(\g_{23}\,\bp_1-\bp_2\,\a_{32}),\label{dm52}\\
\pp_{+}\g_{21}&=&-\g_{21}\,\pp_{+}\phi_+ +\b_{11}-\b_{22}+\sqrt{i}(\a_{23}\,\bp_1-\bp_2\,\a_{31}),\label{dm53}\\
\pp_{+}\g_{23}&=&-\frac{\g_{23}}{2}\pp_{+}({\phi_+-\phi_-})+\b_{13}+\a_{23}-\sqrt{i}\bp_2\,\a_{33}+\sqrt{i}(\a_{22}+\g_{21})\bp_1,\label{dm54}\\
\pp_{+}\g_{31}&=&-\frac{\g_{31}}{2}\pp_{+}({\phi_++\phi_-})-\b_{32}-\a_{31}+\sqrt{i}\a_{33}\,\bp_1-\sqrt{i}\bp_2(\a_{11}+\g_{21}),\label{dm55}\\
\pp_{+}\g_{33}&=&\sqrt{i}(\a_{32}+\g_{31})\bp_1-\sqrt{i}\bp_2(\a_{13}+\g_{23}).\label{dm56}	
\end{eqnarray}

\noindent Note that besides having $\a_{12}=\a_{21}=0$, we find from eqs. (\ref{dm5}), (\ref{dm12}), (\ref{dm45}), (\ref{dm46}) and (\ref{dm50}) that  $\g_{11}=\g_{22} = c_{11}$, $\g_{33} = c_{33}$, $\b_{11}=b_{11} e^{-\phi_-}$, $\b_{22}=b_{11}e^{\phi_-}$, and $\b_{33} =b_{33}$, where the Latin letter $b_{ij}$ and $c_{ij}$ denote arbitrary constants. For convenience, we chose $b_{33}=b_{11}$ and $c_{33}=c_{11}$. Then, substituting in eqs. (\ref{dm10}) and (\ref{dm11}), we get
\br
 \a_{13}+\g_{23}= -(\g_{31}+\a_{32})\,=\, -\sqrt{i}\,c_{11}\bp_-. \label{a57}
\er
In addition, from eqs. (\ref{dm28}), (\ref{dm32}), (\ref{dm48}), and (\ref{dm49}), we find
\br 
 \pa_-(\a_{13}+\a_{32})= 0, \qquad \pa_-(\g_{31}+\g_{23})=0,
\er
and therefore we will consider the simple solution with $\a_{13}=-\a_{ 32}$ and $\g_{31} = -\g_{23}$.
By introducing the following parametrization 
\br 
\b_{12} =b_{12}\,e^{(\phi_+-\l_0)},
\er
and substituting in eqs. (\ref{dm35}) and (\ref{dm41}), we obtain
\br
 \pp_{+}\l_0 &=& - \frac{2b_{11}}{b_{12}}\,e^{-(\ph_+-\l_0)}\sinh\phi_- - \frac{\sqrt{i}}{b_{12}}\,e^{-(\ph_+-\l_0)}(\b_{13}\,\bp_1 -\bp_2\,\b_{32}).\label{a63}\\
 \pp_{-}(\phi_+-\l_0) &=&-\frac{m^{2}c_{11}}{2b_{12}}e^{\l_0}\sinh\phi_- -\frac{\sqrt{i}m}{2b_{12}}e^{-\frac{\phi_+}{2}+\l_0}\a_{13}\left[\p_+\cosh\!\Big({\phi_-\over 2}\Big) +\p_-\sinh\!\Big({\phi_-\over 2}\Big)\],\non\\ \mbox{}\label{a62}
\er
By comparing with eqs. (\ref{f4.45})--(\ref{f4.51}) we find the following constraints on the parameters from the bosonic terms of the above equations, 
\br
\frac{b_{11}}{b_{12}}=\frac{m \s}{4}, \qquad \frac{c_{11}}{b_{12}} =\frac{1}{m \s},\label{a64}
\er
and from the second term in (\ref{a62}) we get,
\br
\a_{13} = -\sqrt{im\s}\, c_{11}\, e^{\frac{(\ph_+-\l_0)}{2}} f_1.
\er
From (\ref{a57}) we get \vspace{-0.3cm}
\br
\g_{23} = -\sqrt{im\s} \,c_{11}\,e^{-\frac{(\ph_+-\l_0)}{2}}\Big[\sinh\Big({\phi_-\over 2}\Big)\tf_1+\cosh\t\, f_1\Big],
\er
and from eqs. (\ref{dm18}) and (\ref{dm19}) and the defect conditions, we find
\br
\a_{31}\,e^{{(\phi_+ +\phi_-)}}+\b_{32}&=& \frac{2\sqrt{i}b_{11}}{\sqrt{m\s}}\, e^{\frac{(\phi_+ +\phi_-) }{2}}\left[\Big(e^{-{\l_0\over 2}}+e^{{\l_0\over 2}}\cosh\t\Big)\tf_1-e^{\frac{\l_0}{2}}\sinh\Big(\frac{\phi_-}{2}\Big)f_1\right],\qquad\quad \mbox{}\label{a67}\\[0.1cm]
\a_{23}\,e^{(\phi_+-\phi_-)}+ \b_{13}&=&\frac{2\sqrt{i}b_{11}}{\sqrt{m\s}}\,e^{\frac{(\phi_+-\phi_-)}{2}}\left[\Big(e^{-{\l_0\over 2}}+e^{{\l_0\over 2}}\cosh\t\Big)\tf_1-e^{\frac{\l_0}{2}}\sinh\Big(\frac{\phi_-}{2}\Big)f_1\right].\label{a68}
\er
A compatible solution for eqs. (\ref{a67}) and (\ref{a68}) is given now by
\br
\a_{23} &=& \a_{31} \,e^{\phi_-},\qquad \quad \a_{31}= \frac{2\sqrt{i}}{\sqrt{m\s}}b_{11}\, e^{-\frac{(\phi_+ -\l_0) }{2}}e^{-\frac{\phi_-}{2}}\left[\cosh\t\, \tf_1-\sinh\Big(\frac{\ph_{-}}{2}\Big)f_1\right],\qquad \mbox{} \\[0.1cm]
\b_{32} &=& \b_{13} \,e^{\phi_-}, \qquad \quad \, \b_{13}=\frac{2\sqrt{i}}{\sqrt{m\s}}b_{11}\, e^{\frac{(\phi_+ -\l_0) }{2}}e^{-\frac{\phi_-}{2}}\tf_1.
\er
Now, by considering eq. (\ref{dm45}), namely
\br
\pa_-\b_{33}&=& \frac{\sqrt{i}m}{2}(\b_{32}e^{-\frac{(\phi_++\phi_-)}{2}}+\a_{31}e^{\frac{(\phi_++\phi_-)}{2}})\p_1+\frac{\sqrt{i}m}{2}\p_2(e^{-\frac{(\phi_+-\phi_-)}{2}}\b_{13}+e^{\frac{(\phi_+-\phi_-)}{2}}\a_{23}),\qquad\,\,\,\mbox{}\non\\
&=&\frac{\si m}{2}\left[\b_{13}e^{-\frac{(\phi_+-\phi_-)}{2}}+e^{\frac{(\phi_+-\phi_-)}{2}}\a_{23}\]\psi_-\,=\,0,\non
\er
the consistency of the above results is verified. In addition, eq. (\ref{a63}) takes the following form
\br
\pa_+\l_0&=& - \frac{2b_{11}}{b_{12}}e^{-(\ph_+-\l_0)}\sinh\phi_--
\frac{2ib_{11}}{b_{12}}\left(1+e^{-(\ph_+-\l_0)}\cosh\t\right)\sinh\Big(\frac{\phi_-}{2}\Big)f_1\tf_1\non\\&&+\frac{2ib_{11}}{b_{12}\sqrt{m\s}}\, e^{-\frac{(\ph_+-\l_0)}{2}}\cosh\Big(\frac{\phi_-}{2}\Big)\bpsi_+\tf_1.\quad \mbox{}
\er
Let us now consider the eqs. (\ref{dm34}) and (\ref{dm51}) involving $\b_{11}$ and $\g_{11}$, namely
\br
\pp_{-}\phi_- &=&-\frac{m^2}{4b_{11}}\g_{21}\,e^{\phi_+}+\frac{m^2b_{12}}{b_{11}}\,e^{-\l_0}+\frac{i}{2}\sqrt{\frac{m}{\s}}\left(e^{\frac{\l_0}{2}}\cosh\t- e^{-\frac{\l_0}{2}}\right)\psi_+\tf_1\non \\&&-\frac{i}{2}\sqrt{\frac{m}{\s}}\, e^{\frac{{\lambda}_{0}}{2}}\sinh\Big(\frac{\ph_{-}}{2}\Big) \psi_+ f_1-\frac{im^2 b_{12}}{4b_{11}}\ e^{\l_0} \cosh\t\cosh\Big(\frac{\phi_-}{2}\Big) f_1\tf_1,\qquad \mbox{}\\
\pp_{+}\phi_- &=&\frac{\g_{21}}{c_{11}}-\frac{b_{12}}{c_{11}}\,e^{(\phi_+-\l_0)}+\frac{i\sqrt{m\s}}{2} \left(e^{\frac{(\ph_{+}-{\lambda}_{0})}{2}}-e^{-\frac{(\ph_{+}-{\lambda}_{0})}{2}}\cosh\t\right) \bpsi_+f_1\non\\&&-\frac{i\sqrt{m\s}}{2}\, e^{-\frac{(\ph_{+}-{\lambda}_{0})}{2}}\sinh\Big(\frac{\ph_{-}}{2}\Big)\bp_{+}\tf_1.\qquad\mbox{}
\er
Then, by comparing with the defect conditions (\ref{f4.45})--(\ref{f4.51}), we find that a suitable solution for $\g_{21}$ is given as follows,
\br
\g_{21} = b_{12}\,e^{-(\phi_+-\l_0)}\left[\sinh^2\Big({\phi_-\over 2}\Big)+\cosh^2\t+i\cosh\t\cosh\Big({\phi_-\over 2}\Big)f_1\tf_1\right].
\er
%
%
In addition, the constraints (\ref{dm9}), (\ref{dm16})--(\ref{dm17}) for $\a_{11}$ and $\a_{22}$ take the following form,
\br 
\a_{11}-\a_{22} &=&\!\!- im\s c_{11}\sinh\Big({\phi_-\over 2}\Big)f_1\tf_1,\\
\a_{11}\, e^{\phi_-} -\a_{22}\, e^{-\phi_-} &=&\,\,  im\s c_{11}\sinh\Big({\phi_-\over 2}\Big)f_1\tf_1,
\er
and then, we get\vspace{-0.4cm}
\br
\a_{22} = \a_{11}\,e^{\phi_-}, \qquad \quad \a_{11}=\frac{i}{2}m\s c_{11}\,e^{-\frac{\phi_-}{2}}f_1\tf_1 . 
\er
Finally, by considering eq. (\ref{dm30}) we find the following relation,
\br
\a_{33} (\psi_+ -\psi_-) = -m\s c_{11}\left[i\cosh\Big({\ph_-\over2}\Big)\tf_1 f_1+\cosh\t\right](\psi_+ -\psi_-).
\er
Now, using that $ \tf_1 f_1\psi_-=0$, we conclude that
\br
\a_{33} = -m\s c_{11}\left[i\cosh\Big({\ph_-\over2}\Big)\tf_1 f_1+\cosh\t\right],
\er
is a consistent solution for the eqs. (\ref{dm26}) and (\ref{dm33}). Therefore, we have found a different solution for the defect matrix, and can be written as follows
\begin{eqnarray}
\cK &=&\left(\begin{array}{c c | c}\cK_{11}&\cK_{12}&\cK_{13}\\[0.2cm] \cK_{21}&\cK_{22}&K_{23} \tabularnewline\hline \mbox{}&\mbox{}&\mbox{}\\[-0.3cm]
\cK_{31}&\cK_{32}&\cK_{33}      \end{array}\right),
\end{eqnarray}
where
\begin{eqnarray}
\cK_{11}&=&\frac{i}{2}m\s c_{11}\,e^{-\frac{\phi_-}{2}}f_1\tf_1+\l^{-1/2}b_{11}\,e^{-\ph_-}+\l^{1/2}c_{11},\\
\cK_{12}&=&\l^{-1/2}b_{12}\,e^{(\ph_+-\l_0)},\\
\cK_{13}&=&-\sqrt{im\s}\, c_{11}\, e^{\frac{(\ph_+-\l_0)}{2}} f_1+\l^{-1/2}\frac{2\sqrt{i}}{\sqrt{m\s}}b_{11}\, e^{\frac{(\phi_+ -\l_0) }{2}}e^{-\frac{\phi_-}{2}}\tf_1,\\
\cK_{21}&=&\l^{1/2}b_{12}\,e^{-(\phi_+-\l_0)}\left[\sinh^2\Big({\phi_-\over 2}\Big)+\cosh^2\t+i\cosh\t\cosh\Big({\phi_-\over 2}\Big)f_1\tf_1\right],\\
\cK_{22}&=&\frac{i}{2}m\s c_{11}\,e^{\frac{\phi_-}{2}}f_1\tf_1+\l^{-1/2}b_{11}\,e^{\ph_-}+\l^{1/2}c_{11}\\
\cK_{23}&=&\frac{2\sqrt{i}}{\sqrt{m\s}}b_{11}\, e^{-\frac{(\phi_+ -\l_0) }{2}}e^{\frac{\phi_-}{2}}\left[\cosh\t\, \tf_1-\sinh\Big(\frac{\ph_{-}}{2}\Big)f_1\right]\non\\&&-\l^{1/2}\sqrt{im\s} \,c_{11}\,e^{-\frac{(\ph_+-\l_0)}{2}}\Big[\sinh\Big({\phi_-\over 2}\Big)\tf_1+\cosh\t\, f_1\Big],\qquad \mbox{}
\er
\br
\cK_{31}&=&\frac{2\sqrt{i}}{\sqrt{m\s}}b_{11}\, e^{-\frac{(\phi_+ -\l_0) }{2}}e^{-\frac{\phi_-}{2}}\left[\cosh\t\, \tf_1-\sinh\Big(\frac{\ph_{-}}{2}\Big)f_1\right]\non\\&&+\l^{1/2}\sqrt{im\s} \,c_{11}\,e^{-\frac{(\ph_+-\l_0)}{2}}\Big[\sinh\Big({\phi_-\over 2}\Big)\tf_1+\cosh\t\, f_1\Big],\\
\cK_{32}&=&\sqrt{im\s}\, c_{11}\, e^{\frac{(\ph_+-\l_0)}{2}} f_1+\l^{-1/2}\frac{2\sqrt{i}}{\sqrt{m\s}}b_{11}\, e^{\frac{(\phi_+ -\l_0) }{2}}e^{\frac{\phi_-}{2}}\tf_1,\\
\cK_{33}&=& -m\s c_{11}\left[i\cosh\Big({\ph_-\over2}\Big)\tf_1 f_1+\cosh\t\right]+\l^{-1/2}b_{11}+\l^{1/2}c_{11},
\end{eqnarray}
where the constants parameters satisfy the following relations,
\br
b_{11} = \frac{m\s}{4}b_{12}, \qquad b_{12} = m\s c_{11}.
\er
\noindent The solution for the differential equations (\ref{gauge}) leads to the following equations,
\begin{eqnarray}
\pa_-(\phi_+ -\l_0) &=& -\frac{m}{2\s}e^{\l_0}\sinh\phi_-  -\frac{im}{2\s}\Big(1+e^{\l_0}\cosh\t\Big) \sinh\Big(\frac{\phi_-}{2}\Big)f_1\tf_1\non\\
&&   -\frac{i}{2}\sqrt{m\over \s}e^{\l_0 \over 2}\cosh\Big(\frac{\phi_-}{2}\Big)\psi_+f_1,\label{bk1}\\
\pa_-\phi_- &=& \frac{m}{\s}\left[e^{-\l_0} - e^{\l_0}\Big(\sinh^2\Big(\frac{\phi_-}{2}\Big) +\cosh^2\tau \Big)\right]\non\\
&&-\frac{i}{2}\sqrt{\frac{m}{\s}}\left[\Big(e^{-{\l_0\over 2}}-e^{{\l_0\over 2}}\cosh\t\Big)\psi_+\tf_1 +e^{{\l_0\over 2}}\sinh\Big(\frac{\phi_-}{2}\Big)\psi_+f_1\right]\non\\
&&-\frac{im}{\s }\,e^{\l_0}\cosh\t \cosh\Big(\frac{\phi_-}{2}\Big)f_1\tf_1,\label{bk2}\\
\pa_+\phi_-&=& m\s\left[ e^{-(\phi_+-\l_0)}\Big(\sinh^2\Big(\frac{\phi_-}{2}\Big) +\cosh^2\tau \Big)-e^{(\phi_+-\l_0)} \right]\non\\
&& +\frac{i\sqrt{m\s}}{2}\Big(e^{\frac{\left(\phi_+-\l_0\)}{2}}-e^{-\frac{\left(\phi_+-\l_0\)}{2}}\cosh\tau\Big)\bpsi_+f_1 \non \\
&&-\frac{i\sqrt{m\s}}{2}\,e^{-\frac{\left(\phi_+-\l_0\)}{2}}\sinh\Big(\frac{\phi_-}{2}\Big)\bpsi_+\tf_1\non\\
&&+im\s \,e^{-(\phi_+-\l_0)}\cosh\t \cosh\Big(\frac{\phi_-}{2}\Big)f_1\tf_1,\label{bk3}\\
\pa_+\l_0 &=& -\frac{m\s}{2}\,e^{-(\phi_+-\l_0)}\sinh\phi_-  -\frac{im\s}{2}\Big(1+e^{-(\phi_+-\l_0)} \cosh\t\Big)\sinh\Big(\frac{\phi_-}{2}\Big) f_1\tf_1\non\\
&&  +\frac{i\sqrt{m\s}}{2}\, e^{-\frac{\left(\phi_+-\l_0\)}{2}}\cosh\Big(\frac{\phi_-}{2}\Big)\bpsi_+\tf_1,\label{bk4}\\
\p_- &=&\sqrt{\frac{m}{\s}}\left[e^{\frac{\l_0}{2}}\sinh\Big(\frac{\phi_-}{2}\Big)f_1-\Big(e^{-{\l_0\over 2}}+e^{{\l_0\over 2}}\cosh\t\Big)\tf_1\right],\label{bkp-}\\
\bp_-&=& \sqrt{m\s}\!\left[\Big(e^{\frac{\left(\phi_+-\l_0\)}{2}}\!+e^{-\frac{\left(\phi_+-\l_0\)}{2}}\cosh\tau\Big)f_1 +e^{-\frac{\left(\phi_+-\l_0\)}{2}}\sinh\Big(\frac{\phi_-}{2}\Big)\tf_1\right],\qquad \quad\mbox{}\label{bkbp-}
\er
\br
\pa_+\tf_1  &=&  -\frac{\sqrt{m\s}}{2}e^{-\frac{\left(\phi_+-\l_0\)}{2}}\sinh\Big(\frac{\phi_-}{2}\Big)\bpsi_+  +\frac{m\s}{2}\cosh\Big(\frac{\phi_-}{2}\Big)f_1\\
&& +\frac{m\s}{2} e^{-(\phi_+-\l_0)}\cosh \tau\,\cosh\Big(\frac{\phi_-}{2}\Big)f_1 \label{bk7}\\
\pa_-\tf_1  &=& \frac{1}{2}\sqrt{\frac{m}{\s}}\Big(e^{-{\l_0\over 2}}+e^{{\l_0\over 2}}\cosh\t\Big)\psi_+-\frac{m}{2\s} \left[1+ e^{\l_0}\cosh \tau\right]\!\cosh\Big(\frac{\phi_-}{2}\Big)f_1,\label{bk8}\\
\pa_- f_1 &=& -\frac{1}{2}\sqrt{\frac{m}{\s}}
e^{{\l_0\over 2}}\sinh\Big(\frac{\phi_-}{2}\Big)\psi_++\frac{m}{2\s} \left[1 + e^{\l_0}\cosh \tau\right]\!\cosh\Big(\frac{\phi_-}{2}\Big)\tf_1,\label{bk5}\\
\pa_+ f_1 &=&  -\frac{\sqrt{m\s}}{2}\Big(e^{\frac{\left(\phi_+-\l_0\)}{2}}+e^{-\frac{\left(\phi_+-\l_0\)}{2}}\cosh\tau\Big)\bpsi_+ -\frac{m\s}{2} \cosh\Big(\frac{\phi_-}{2}\Big)\tf_1\non\\
&& -\frac{m\s}{2}  e^{-(\phi_+-\l_0)}\cosh \tau\,\cosh\Big(\frac{\phi_-}{2}\Big)\tf_1,\label{bk6}
\end{eqnarray}
which correspond to the type-II super-B\"acklund transformation for the $N=1$ sshG model with two arbitrary parameters $(\s, \t)$ derived through the fusing defects in section \ref{fusing}. Consistency with the super-B\"acklund  presented in section \ref{backlund} is shown by parametrizing $\tau=i\pi/2$, and
\br
 \om_1=-\om_4 =-\sqrt{\frac{m}{\s}},\qquad  \om_2 =-\om_3 =\sqrt{m\s}.
\er
Finally, we note that this form of the type-II super-B\"acklund transformation can  also be written in terms of superfields in the following way,
\br
D_{-}(\Phi_{+}-{\Lambda}) &=& -\sqrt{\frac{m}{\s•}}\ e^{\frac{{\Lambda}}{2}}\,f\,\cosh\Big(\frac{\Phi_{-}}{2}\Big),\\
D_{+}{\Lambda} &=&\sqrt{m\s}\ e^{-\frac{\left(\Phi_{+}-{\Lambda}\right)}{2}}\,\tilde{f}\,\cosh\Big(\frac{\Phi_{-}}{2}\Big),\\
D_{+}\Phi_{-} &=&-\sqrt{m\s} \left[\( e^{\frac{(\Phi_{+}-{\Lambda})}{2}} +\cosh\tau \, e^{-\frac{\left(\Phi_{+}-{\Lambda}\right)}{2}}\)f
+ e^{-\frac{\left(\Phi_{+}-{\Lambda}\right)}{2}}\sinh\Big(\frac{\Phi_{-}}{2}\Big)\tf \right],\qquad\quad  \mbox{}\\
D_{-}\Phi_{-} &=&\sqrt{\frac{m}{\s•}}\left[ \( e^{-\frac{{\Lambda}}{2}}+\cosh\tau\, e^{\frac{{\Lambda}}{2}}\)\tilde{f} -e^{\frac{{\Lambda}}{2}}\sinh\Big(\frac{\Phi_{-}}{2}\Big)f\right],
\er
where the auxiliary fields satisfy the relations,
\begin{eqnarray}
D_{+}f &=&-i\sqrt{m\s}\left[ e^{\frac{(\Phi_{+}-{\Lambda)}}{2}}-\cosh\tau\, e^{-\frac{(\Phi_{+}-{\Lambda)}}{2}}\right],\qquad 
D_{-}f=i\sqrt{\frac{m}{\s}}\ e^{\frac{{\Lambda}}{2}}\,\sinh\Big(\frac{\Phi_{-}}{2}\Big),\qquad \quad\quad  \mbox{}\\[0.1cm]
D_{-}\tilde{f} &=& i\sqrt{\frac{m}{\s}}\left[ e^{-\frac{{\Lambda}}{2}}-\cosh\tau\,  e^{\frac{{\Lambda}}{2}}\right], \qquad \qquad D_{+}\tilde{f}=i\sqrt{m\s}\ e^{-\frac{\left(\Phi_{+}-{\Lambda}\right)}{2}}\ \sinh\Big(\frac{\Phi_{-}}{2}\Big), \quad \mbox{}
\end{eqnarray}
and can also  be split in components by using expansions of the superfields $\L$, $f$, and $\tf$ given in  (\ref{B1.13})--(\ref{B1.15}). 



\newpage
\begin{thebibliography}{99}

\bibitem{Bow1} P. Bowcock, E. Corrigan and C. Zambon, \emph{Classically integrable field theories with defects}, {\it Int.\ J.\ Mod.\ Phys.} \textbf{A 19S2} (2004) 82 [{\tt hep-th/0305022}].

\bibitem{Bow2} P. Bowcock, E. Corrigan and C. Zambon, \emph{Affine Toda field theories with defects},  {\it JHEP} {\bf 01} (2004) 056 [{\tt hep-th/0401020}].

\bibitem{Bow3} P. Bowcock, E. Corrigan and C. Zambon, \emph{Some aspects of jump-defects in the quantum sine-Gordon model}, {\it JHEP} {\bf 08} (2005) 23 [{\tt arXiv:0506169}].

\bibitem{Corr1} E. Corrigan and C. Zambon, \emph{Jump-defects in the nonlinear Schr\"odinger model and other non-relativistic field theories}, {\it Nonlinearity} {\bf 19} (2006) 1447 [{\tt nlin/0512038}].

\bibitem{FLZ1} J.F. Gomes, L.H. Ymai, and A.H. Zimerman, \emph{Classical integrable super sinh-Gordon equation with defects}, {\it J. Phys.} \textbf{A 39} (2006) 7471 [{\tt hep-th/0601014}]. 

\bibitem{Corr2} E. Corrigan and C. Zambon, \emph{On purely transmitting defects in affine Toda field theory}, {\it JHEP} {\bf 07} (2007) 001 [{\tt arXiv:0705.1066}].

\bibitem{FLZ2}  J.F. Gomes, L.H. Ymai and A.H. Zimerman, \emph{Integrability of a classical $N= 2$ super sinh-Gordon model with jump defects}, {\it JHEP} {\bf 03} (2008) 001  [{\tt hep-th/0710.1391}].

\bibitem{Corr3} 	E. Corrigan and C. Zambon, \emph{Comments on defects in the a(r) Toda field theories}, {\it J. Phys.} {\bf A 42} (2009) 304008 [{\tt arXiv:0902.1307}].

\bibitem{Ale} A.R. Aguirre, J.F. Gomes, L.H. Ymai and A.H. Zimerman,
  \emph{Thirring model with jump defect},
  {\tt PoS(ISFTG)031}
  [{\tt nlin/0910.2888}].

\bibitem{Corr4} E. Corrigan  and C. Zambon, \emph{A new class of integrable defects},  {\it J. Phys.} {\bf A 42} (2009) 475203 [{\tt hep-th/0908.3126}].

\bibitem{Corr5} E. Corrigan and C. Zambon, \emph{Integrable defects in affine Toda field theory and infinite dimensional representations of quantum groups}, {\it Nucl. Phys} {\bf B 848} (2011) 545 [{\tt arXiv:1012.4186}].


\bibitem{Corr10} E. Corrigan and C. Zambon, \emph{A transmission matrix for a fused pair of integrable defects in the sine-Gordon model}, {\it J. Phys.} {\bf A 43}  (2010) 345201 [{\tt arXiv:1006.0939}].


\bibitem{CR} C. Robertson, \emph{Folding defect affine Toda field theories}, {\it J. Phys.} {\bf A 47} (2014) 185201 [{\tt arXiv:1304.3129}].


\bibitem{Cau1} 
  V.~Caudrelier,
  \emph{On a systematic approach to defects in classical integrable field theories},
 {\it Int.\ J.\ Geom.\ Meth.\ Mod.\ Phys.}\  {\bf 5} (2008) 1085 [{\tt arXiv:0704.2326v2}]


\bibitem{Ale2} A.R. Aguirre, J.F. Gomes, L.H. Ymai and A.H. Zimerman, \emph{Grassmanian and bosonic Thirring models with jump defects}, {\it JHEP} {\bf 02} (2011) 017 [{\tt nlin/1012.1537}].

\bibitem{Ale3} A.R. Aguirre, T.R. Araujo, J.F. Gomes, and A.H. Zimerman, \emph{Type-II B\"acklund transformations via gauge transformations}, {\it JHEP} {\bf 12} (2011) 056 [{\tt nlin/1110.1589}].

\bibitem{Ale4} A.R. Aguirre, \emph{Inverse scattering approach for massive Thirring models with integrable type-II defects}, {\it J. Phys.} {\bf A 45} (2012) 205205  [{\tt arXiv:1111.5249}].

\bibitem{Kundu} I. Habibullin and A. Kundu, \emph{Quantum and classical integrable sine-Gordon model with defect}, {\it Nucl.\ Phys.} {\bf B 795} (2008) 549 [{\tt hep-th/0709.4611}].

\bibitem{Doi1}  J. Avan and A. Doikou, \emph{Liouville integrable defects: the nonlinear Schr\"odinger paradigm}, {\it JHEP} {\bf 01} (2012) 040  [arXiv:1110.4728].

\bibitem{Doi2} J. Avan and A. Doikou, \emph{The sine-Gordon model with integrable defects revisited}, {\it JHEP} {\bf11} (2012) 008 [{\tt hep-th/1205.1661}].

\bibitem{Doi3} A. Doikou and N. Karaiskos,  \emph{Sigma models in the presence of dynamical point-like defects}, {\it Nucl. Phys.}  {\bf B 867} (2013) 872 [{\tt arXiv:1207.5503}].

\bibitem{Doi4} A. Doikou, \emph{A note on $gl_N$  type-I integrable defects}, {\it J. Stat. Mech.} (2014) P02002 [{\tt  arXiv:1308.1790}].

\bibitem{Doi5} A. Doikou, \emph{Classical impurities associated to high rank algebras}, {\it
Nucl. Phys.} {\bf B 884} (2014) 142 [{\tt arXiv:1312.4786}].

\bibitem{Doi6} A. Doikou, \emph{Jumps and twists in affine Toda field theories}, {\it Nucl. Phys.} {\bf B 893} (2015) 107 [{\tt arXiv:1404.7329}].

\bibitem{Cau2} V. Caudrelier and A. Kundu, \emph{A multisymplectic approach to defects in integrable classical field theory}, {\it JHEP} {\bf 02} (2015) 088 [{\tt arXiv:1411.0418}].

\bibitem{Cau3} V. Caudrelier, \emph{Multisymplectic approach to  integrable defects in the sine-Gordon model}, {\it J. Phys.} {\bf A 48} (2015) 195203 [{\tt arXiv:1411.0418}].

\bibitem{Nathaly1} A.R. Aguirre, J.F. Gomes, N.I. Spano, A.H. Zimerman, \emph{$N=1$ super sinh-Gordon model with defects revisited}, {\it JHEP} {\bf 02} (2015) 175 [{\tt arXiv:1412.2579}]


\bibitem{Ale6} A.R. Aguirre, \emph{Type-II defects in the super-Liouville theory},
  {\it J.\ Phys.\ Conf.\ Ser.}  {\bf 474} (2013)  012001 [arXiv:1312.3463].
  

\bibitem{ChaiKul} M. Chaichian and P. Kulish, \emph{On the method of inverse scattering problem and B\"acklund transformations for supersymmetric equations}, {\it Phys. Lett} \textbf{B 78} (1978) 413.

  

\bibitem{Zambon1} E. Corrigan and C. Zambon, \emph{Infinite dimension reflection matrices in the sine-Gordon model with a boundary}, {\it JHEP} {\bf 06} (2012) 050 [{\tt arXiv:1202.6016}].


\bibitem{Ale5}  A.R.~Aguirre, J.F.~Gomes, L.H.~Ymai and A.H.~Zimerman,
  \emph{N=1 super sinh-Gordon model in the half line: Breather solutions},
  {\it JHEP} {\bf 04}  (2013) 136
  [{\tt arXiv:1304.4582}].
  
  
 \bibitem{Zambon2} C. Zambon, \emph{The classical nonlinear Schr\"odinger model with a new integrable boundary}, {\it JHEP} {\bf 08} (2014) 036 [{\tt arXiv:1405.0967}].
  
  
  
\end{thebibliography}
\end{document}